\documentclass[journal]{IEEEtran}
\usepackage{setspace}
\usepackage{cite}
\ifCLASSINFOpdf
   \usepackage[pdftex]{graphicx}
   \graphicspath{{../pdf/}{../jpeg/}}
   \DeclareGraphicsExtensions{.pdf,.jpeg,.png}
\else
   \usepackage[dvips]{graphicx}
   \graphicspath{{../eps/}}
   \DeclareGraphicsExtensions{.eps}
\fi
\usepackage{epstopdf}
\usepackage{stfloats}
\usepackage{amsthm}
\usepackage{color}
\usepackage{amsmath}
\usepackage{bm}

\usepackage{subfigure}
\usepackage{caption}
\usepackage{algorithm}
\usepackage{algorithmic}
\usepackage{amssymb}

\begin{document}
\captionsetup[figure]{labelfont={default},labelformat={default},labelsep=period,name={Fig.}}
\title{Joint MIMO Transceiver and Reflector Design for Reconfigurable Intelligent Surface-Assisted Communication}
\author{Yaqiong Zhao, \emph{Student Member, IEEE,}
Jindan Xu, \emph{Member, IEEE,}
Wei Xu, \emph{Senior Member, IEEE,}\\
Kezhi Wang, \emph{Senior Member, IEEE,}
Xinquan Ye,
Chau Yuen, \emph{Fellow, IEEE,}
and Xiaohu You, \emph{Fellow, IEEE}
%<-this % stops a space
}

\maketitle

\begin{abstract}
In this paper, we consider a reconfigurable intelligent surface (RIS)-assisted multiple-input multiple-output communication system with multiple antennas at both the base station (BS) and the user. We plan to maximize the achievable rate through jointly optimizing the transmit precoding matrix, the receive combining matrix,
and the RIS reflection matrix under the constraints of the transmit power at the BS and the unit-modulus reflection at the RIS. Regarding the non-trivial problem form, we initially reformulate it into an considerable problem to make it tractable by utilizing the relationship between the achievable rate and the weighted minimum mean squared error. Next, the transmit precoding matrix, the receive combining matrix, and the RIS reflection matrix are alternately optimized. In particular, the optimal transmit precoding matrix and receive combining matrix are obtained in closed forms. Furthermore, a pair of computationally efficient methods are proposed for the RIS reflection matrix, namely the semi-definite relaxation (SDR) method and the successive closed form (SCF) method. We theoretically prove that both methods are ensured to converge, and the SCF-based algorithm is able to converges to a Karush-Kuhn-Tucker point of the problem.

\end{abstract}


% \fi
\IEEEpeerreviewmaketitle
\vspace{-5.pt}
\section{Introduction}
Millimeter-wave (mmWave) communication plays a pivotal role  in the next generation of wireless communication systems since it enables the remarkable increase in data
rates by utilizing the spectrum recourses between 30 GHz and
300 GHz \cite{zhao2021joint,rappaport2013millimeter,gao2022integrated}. Acting as another key technology in future wireless communication, massive multiple-input multiple-output (MIMO)
is capable of providing the large array gain to compensate
the severe path loss that mmWave signals undergo.
However, one of the remaining issues is that mmWave signals
are highly sensitive to blockage, which leads to channel
sparsity and low-rank channel structures \cite{yu2018hardware,wong2017key,xu2023toward,he2024unlocking}.

In order to alleviate the negative effects caused by blockage, reconfigurable intelligent surface (RIS) was advocated as a pioneering technology to establish favorable wireless channels by creating a virtual line-of-sight (LoS) link to bypass the obstacle \cite{basar2019wireless, gacanin2020wireless,alexandropoulos2021reconfigurable,xu2023reconfiguring}. RIS is an artificial metal-surface comprising hundreds or even thousands of low-cost passive reflective components. Specifically, each component of RIS is capable of introducing a specified phase shift to
the incident electromagnetic wave with the help of the dedicated controller \cite{wang2020energy,xu2023edge,he2021low}. Without regard to the practical
implementation, what makes RIS widely concerned is that it can amplify and forward the impinging signal without additional power. In comparison to the classic amplify
and forward (AF) relay, RIS neither produces additional thermal noise nor consumes extra power since there is no active radio frequency (RF) component needed \cite{chen2023relay,li2022interplay}. Therefore, RIS is actually a cost-effective and energy-efficient device that perfectly matches the green and sustainable communication trend. Moreover, RISs can be easily and vastly installed on building
surfaces to realize enhanced signal coverage, due to their attractive advantages
of light weight and small sizes. Note that the terms large intelligent surfaces and intelligent reflecting surfaces also revolve around the concept of RIS but emphasizes scale and the reflective properties of these surfaces, respectively \cite{ma2024active,elmossallamy2020reconfigurable}.

Among the numerous merits of RIS, one most attractive thing is its ability in achievable rate improvement. By judiciously tuning the phase shifts in accordance with the instantaneous wireless channels, the electromagnetic signals transmitted by the RIS can be received coherently with those via other propagation paths to strengthen the received signal power. As a result, considerable performance gain can be harvested by a smartly designed RIS reflection. Specifically, both theoretical analysis and numerical simulations have validated that a properly designed RIS with $N$ reflecting elements is anticipated to provide a beamforming gain of ${\cal O}(N^2)$ while the traditional multiple-antenna technology is only able to provide a beamforming gain of ${\cal O}(N)$ \cite{wu2019intelligent,wu2019beamforming}. On the one side, enlarging $N$ renders an increased number of reflecting elements to harvest the associated wireless energy transmitted from the BS, which yields a receive antenna gain of ${\cal O}(N)$. Conversely, the reflect beamforming towards the user (UE) by the RIS provides another transmit beamforming gain of ${\cal O}(N)$.

Against the above background, existing works have investigated the spectrum efficient design in RIS-assisted systems. Generally, the primary challenge in dealing with such problems arises from the non-trivial unit-modulus constraints imposed on the RIS reflection coefficients. Although previous studies \cite{han2019large,guo2019weighted,li2020weighted} have addressed such constraints using diverse approaches like semidefinite relaxation (SDR), they primarily focused on the single-antenna systems such as single-input single-output (SISO) and multiple-input single-output (MISO) communication, and only few works looked into RIS-assisted multi-input multi-output (MIMO) communication due to the more sophisticated parameter optimization \cite{feng2021passive,ning2020beamforming,zhang2020capacity}. For example, \cite{feng2021passive} and \cite{ning2020beamforming} indirectly solved the capacity maximization problem for a RIS-assisted point-to-point MIMO system through changing the objective function from the capacity to the received signal-to-noise (SNR). By studying the specific structure of the channel capacity expression, \cite{zhang2020capacity} developed an element-wise based optimization algorithm for capacity maximization through alternatively obtaining one of the reflection coefficients or the transmit precoder while keeping other optimization variables fixed. However, the element-wise optimization involves much processor computation time and no optimality is guaranteed.

Aiming at maximizing the achievable rate for a RIS-assisted massive MIMO system, we develop effective solutions for the corresponding parameter design in this paper. More specifically, the main contributions of this paper are listed as follows\footnote{In this paper, we mainly investigate the passive RIS-assisted wireless communications. However, the proposed algorithm can also be regarded as a more complicated version of its
active counterpart and provides inspiration for future researchers. The trade-off between the performance of active and passive RISs in terms of the number
of RIS elements and the additional power consumption have also been investigated in \cite{zhang2022active}.}:
\begin{itemize}
  \item  First, we establish the achievable rate maximization problem for a point-to-point MIMO system assisted by RIS, which is hard to solve with the nonconvex objective function and the unit-modulus constraint. Regarding this sophisticated problem, we initially convert it to an considerable form by utilizing the relationship between the achievable rate and the weighted minimum mean-squared error (WMMSE). This allows us to alternatively optimize the RIS reflection matrix and the transceiver in a more efficient way.
  \item With an arbitrary RIS reflection matrix, the optimal transmit precoding matrix and the receive combining matrix are obtained in closed forms by exploiting the Lagrangian multiplier approach. After a series mathematical calculations and transformations, the original RIS reflection matrix design problem is converted to a non-convex quadratically constrained quadratic program (QCQP) with the unit-modulus constraint. Then for the nonconvex QCQP problem, we propose two computationally-efficient optimization algorithms on the basis of the semi-definite relaxation (SDR) and successive closed forms (SCF) methods. Specifically, the SCF-based algorithm allows a semi-closed form solution to the RIS reflection matrix. Moreover, we theoretically prove that both algorithms are guaranteed to converge, and the SCF-based algorithm converges to a Karush-Kuhn-Tucker (KKT) point of the equivalent problem.
  \item Finally, we provide numerical results to validate the satisfactory performance of the proposed algorithms. Particularly, it is demonstrated that with a judiciously designed RIS phase shifts, various key performance metrics of the RIS-aided MIMO system can be greatly improved, such as achievable rate, normalized mean squared error (NMSE), channel total power. Moreover, we extend our designs to more practical situations, i.e., discrete-phase cases, by adopting a quantized phase projection approach. It is observed that even 2 bits of quantization phase control may work well with very limited performance degradation.
\end{itemize}

The remainder of this paper is listed as follows. System model is introduced in Section \uppercase\expandafter{\romannumeral2}, followed by problem formulation part. In Section \uppercase\expandafter{\romannumeral3}, the WMMSE-based problem is reformulated at first, and then the optimal transceiver is presented. We then
propose two alternating minimization algorithms to solve the RIS reflection design problem in Section \uppercase\expandafter{\romannumeral4}, with the optimality and convergence analysis of the proposed algorithms provided as well. Numerical simulation results are provided in Section \uppercase\expandafter{\romannumeral5} to show the advantage of our proposed design, and the conclusion is summarized in Section \uppercase\expandafter{\romannumeral6}.

\emph{Notations}: Operators diag$\left(  \cdot  \right)$ and Re$\{\cdot\}$ are utilized to denote a diagonal matrix and the real component of a complex value. vec$(\cdot)$ means vectorization. The symbol $\otimes$ returns the Kronecker product. The notation ${\bf A}^{1/2}$ is the squared root of matrix ${\bf A}$ by Cholesky decomposition and $\otimes$ means the Kronecker product. The notations tr$\{\cdot\}$ and $\mathbb{E}\{\cdot\}$ respectively stand for the trace operation and the expectation operation. Notation $\|\cdot\|_1$, $\|\cdot\|_2$, $\|\cdot\|_{F}$ and $\|\cdot\|_{\infty}$ respectively, denote the $L_1$ norm, the $L_2$ norm, the Frobenius norm and the infinite norm of the input matrix. $\lambda_{\text{max}}(\cdot)$ corresponds to the maximum eigenvalue. ${\cal {CN}}(0,\sigma^2)$ models the standard complex Gaussian distribution whose mean is zero and variance is $\sigma^2$. ${\bf I}_m$ is the $m \times m$ identity matrix. $\bf 1$ and $\bf 0$ means an all-one and all-zero vectors, respectively. Finally, arg$\{\cdot\}$ stands for the phase of a complex value and ${\cal O}$ denotes the standard big-O notation.
\section{System Model and Problem Formulation}
\begin{figure}[!t]
\setlength{\abovecaptionskip}{0pt}
\setlength{\belowcaptionskip}{-10pt}
\centering %使得插入的照片居中显示
\includegraphics[width=8.5cm,height=5cm]{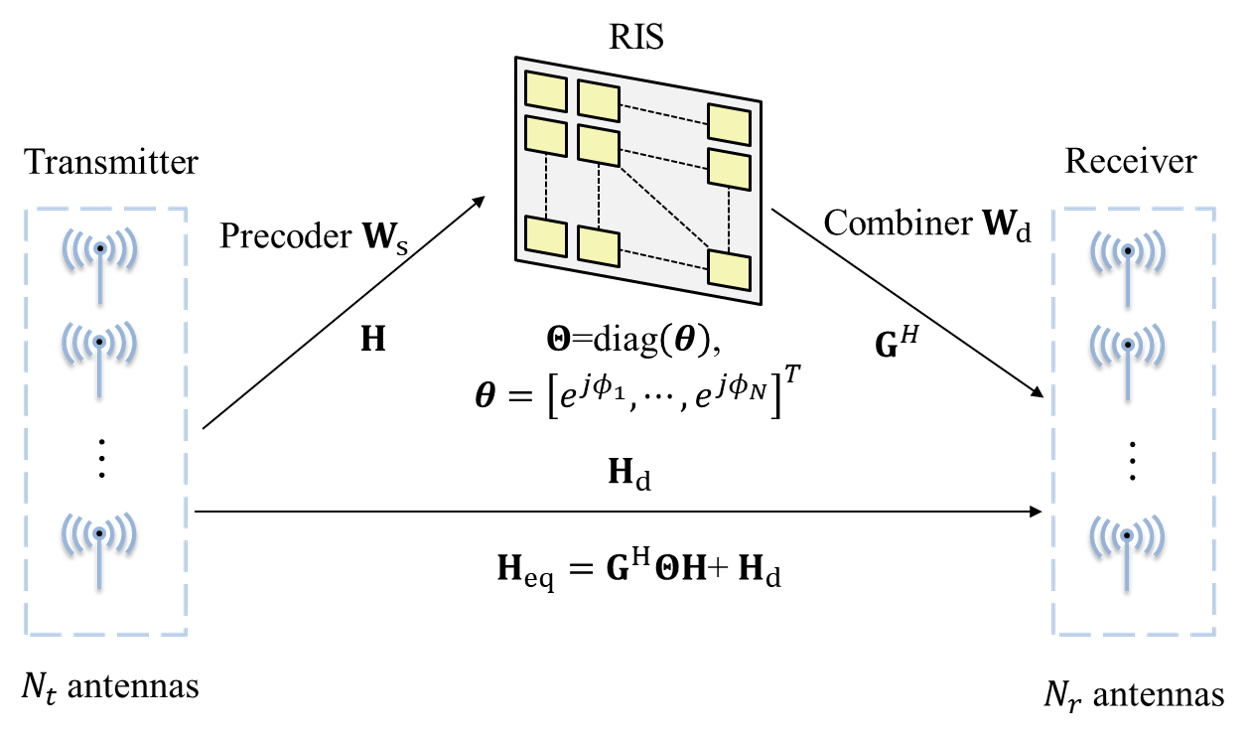} % 图片标题
\caption{\small{System model for a RIS-assisted MIMO system.}}
\label{fig:0}     %
\end{figure}
\subsection{System Model}
We consider a downlink point-to-point MIMO system as depicted in Fig. 1. More specifically, the BS equipped with $N_t$ transmit antennas communicates $N_s$ data streams to a UE with $N_r$ receive antennas through a RIS with $N$ passive reflective components. Taking into account the severe path attenuation at high
frequency, the transmit signals that are passively transmitted via RIS greater than once are neglected. Since the received signal is a summation of both the BS-UE (direct) link and RIS-UE link, the system performance can be significantly improved compared with that without RIS.

For convenience, quasi-static flat-fading is assumed for all channels within our studied configuration. We denote ${\bf H}_d\in {\mathbb C}^{N_r\times N_t}$, ${\bf G}\in {\mathbb C}^{N\times N_r}$ and ${\bf H}\in {\mathbb C}^{N\times N_t}$ as the channels of the BS-UE link, the UE-RIS link, and the BS-RIS link, respectively. The reflection matrix of the RIS is ${\bm\Theta}= \text {diag}\left({\bm\theta}\right)\in {\mathbb C}^{N\times N}$ where ${\bm\theta}= \left[e^{\jmath\varphi_1},\cdots,e^{\jmath\varphi_N}\right]^T $, in which $\varphi_n\in[0,2\pi)$ stands for the phase shift of the $n$th reflecting element. Note that in writing this expression, we have implicitly assumed that the reflection mode of RIS is full reflection, i.e., there is no
energy loss. In practice, it is difficult to realize full reflection due to hardware impairments. However, if we adopt the same energy loss factors for all the reflecting elements, then the proposed two algorithms in the next section still apply since the constant-modulus constraint does not change. For the more complicated scenario of different energy loss factors for different reflecting elements, since this paper mainly focus on the joint design of the active transceiver and passive RIS beamforming, it will be left for further research due to page limit. Filtered by the baseband precoding matrix ${\bf W}_s\in{\mathbb C}^{N_t\times N_s}$, the signal arrived at the UE can be expressed as
\begin{equation}\label{eq:system}
{\bf y}_d = {\bf G}^H{\bm\Theta}{\bf H}{\bf W}_s{\bf s} + {{\bf{H}}_d}{{\bf{W}}_s}{\bf{s}} + {\bf{n}},
\end{equation}
in which ${\bf s}\in{\mathbb C}^{N_s\times 1}$ denotes the transmit signal vector with normalized power $\mathbb{E}\{{\bf ss}^H\}={\bf I}_{N_s}$, and ${\bf n}\sim{\cal {CN}}\left({\bf 0},\sigma^2{\bf I}_{N_r}\right)$ models the complex Gaussian noise with noise power $\sigma^2$. Note that the transmit power satisfies tr$\left({\bf W}_s{\bf W}_s^H\right)\leq P$ where $P$ is the total power available.

Then, a receive combining matrix ${\bf W}_d\in{\mathbb C}^{N_r\times N_s}$ is implemented at the UE side, so the processed signal after receive combining can be expressed as
\begin{equation}\label{eq:combine}
{\bf y} \!=\!{\bf W}_d^H{\bf y}_d\!=\!{\bf W}_d^H{{\bf{G}}^H}{{\bm\Theta}{\bf H}}{{\bf{W}}_s}{\bf{s}} + {\bf W}_d^H{{\bf{H}}_d}{{\bf{W}}_s}{\bf{s}} + {\bf W}_d^H{\bf n}.
\end{equation}
\subsection{Problem Formulation}
When Gaussian symbols are transmitted, the achievable rate of the considered system is thus given by
\begin{small}
\begin{equation}\label{eq:rate}
R=\text{log}_2\left(\left|{\bf I}_{N_s} + \frac{1}{\sigma ^2}\left({\bf W}_d^H{\bf W}_d\right)^{-1}{\bf W}_d^H{\bf H}_{\text{eq}}{\bf W}_s{\bf W}_s^H{\bf H}_{\text{eq}}^H{\bf W}_d\right|\right),
\end{equation}
\end{small}where ${\bf H}_{\text{eq}}\triangleq {\bf G}^H{\bm\Theta}{\bf H}+{\bf H}_d$ denotes the equivalent channel from the BS to the UE which includes both the direct link and the reflection link through RIS.

Compared with the classic rate expression of traditional MIMO system in \cite{goldsmith2003capacity}, the achievable rate of the considered system is not only dependent on the transmit precoder and the receiver combiner, but also on the RIS phase shift, since it can influence the equivalent channel matrix ${\bf H}_{\text{eq}}$ and the corresponding optimal transceiver. Generally, since RISs are mostly passive reflecting devices without any signal processing abilities, the RIS reflection coefficients are appointed by the BS through the dedicated RIS controller according to different performance goals.

Motivated by the above discussions, we aim at maximizing the achievable rate of the above system through a joint design of the transmit precoding matrix ${\bf W}_{\text s}$, the RIS reflection matrix $\bm\Theta$, and the receive combining matrix ${\bf W}_d$, with the total power constraint at the BS and the unit-modular constraints on the reflection coefficients. The associated problem is then expressed as
\begin{subequations}
\begin{align}
({\cal P})\quad&{\mathop{\text{max}}\limits_{_{{\bm\Theta},{\bf W}_s,{\bf W}_d}}}\quad R\label{eq:problema}\\
&\quad\text{s.t.}~\quad\quad \text {tr}\left({\bf W}_s{\bf W}_s^H\right)\leq P,\label{eq:problemb}\\
&\quad\quad\quad\quad~{\bm\Theta}={\text{diag}}({\bm\theta}),\label{eq:problemc}\\
&\quad\quad\quad\quad~|\theta_n|=1,\forall n\in\{1,2,\cdots,N\}\label{eq:problemd}
\end{align}
\end{subequations}
where $\theta_n=e^{\jmath\varphi_n}$ is the $n$th element of $\bm\theta$.

For simplicity, we assume all involved channels are available at both the BS and the UE, i.e., ${\bf H}_d$, $\bf G$, and $\bf H$. which can be obtained by some existing advanced channel estimation techniques such as cascaded channel estimation followed by a matrix factorization \cite{alexandropoulos2017position,he2021low}.

Note that problem $(\cal P)$ is shown to be non-convex mainly because the achievable rate is non-concave with regard to RIS reflection matrix $\bm\Theta$, not to mention the nonconvex unit-modulus constraint in (\ref{eq:problemd}). In addition, the coupling effect between the transmit precoding matrix ${\bf W}_s$, the receive combining matrix ${\bf W}_d$, and the RIS reflection matrix $\bm\Theta$ further aggravates the challenge. As countermeasures,
two efficient methods are advocated in the following to deal with problem $(\cal P)$.
\section{Low-Complexity Algorithm Design}
This section studies the joint design of the transceiver and the RIS reflector to realize achievable rate maximization of the considered system. To this end, first we reformulate the original problem design and transform the rate maximization problem into the WMMSE problem. Then, we propose efficient algorithms to iteratively and alternatively solve the optimization variables. The objective value will be improved after each iteration step, and finally converges.
\subsection{Problem Reformulation}
In the sequel, we resort to the classic relationship between the achievable rate and the MSE to reformulate problem $\cal P$. In particular, the MSE matrix of the received signal after combining is calculated as
\begin{small}
\begin{align}\label{eq:MSE}
{\bf E} &\!=\!\mathbb{E}\left\{({\bf s}-{\bf y})({\bf s}-{\bf y})^H \right\}\nonumber\\
&\!=\!\left({\bf I}_{N_s}-{\bf W}_d^H{\bf H}_{\text{eq}}{\bf W}_s\right)\left({\bf I}_{N_s}-{\bf W}_d^H{\bf H}_{\text{eq}}{\bf W}_s\right)^H+{\sigma^2}{\bf W}_d^H{\bf W}_d,
\end{align}
\end{small}Considering that the optimal solution to ${\bf W}_d$ is the well-known MMSE combiner:
\begin{equation}\label{eq:Wd}
 {\bf W}_d= \left({\sigma^2}{\bf I}_{N_r}+{\bf H}_{\text{eq}}{\bf W}_s{\bf W}_s^H{\bf H}_{\text{eq}}^H\right)^{-1}{\bf H}_{\text{eq}}{\bf W}_s.
\end{equation}
By substituting (\ref{eq:Wd}) into (\ref{eq:MSE}), it yields
\begin{equation}\label{eq:MSE1}
{\bf E} = {\bf I}_{N_s}-{\bf W}_s^H{\bf H}_{\text{eq}}^H\left({\sigma^2}{\bf I}_{N_r}+{\bf H}_{\text{eq}}{\bf W}_s{\bf W}_s^H{\bf H}_{\text{eq}}^H\right)^{-1}{\bf H}_{\text{eq}}{\bf W}_s.
\end{equation}
To proceed further, we invoke the following lemma to convert the original sophisticated optimization problem into an easier one.

\emph{Lemma 1:} By introducing an auxiliary weight matrix ${\bf W}\succeq {\bf 0}$, the original problem $(\cal P)$ is equivalent to
\begin{align}\label{eq:p1}
({\cal P}_1)\quad&{\mathop{\text{min}}\limits_{_{{\bf W}_s,{\bf W}_d,{\bf W},{\bm\Theta}}}}\quad \text{tr}\{{\bf WE}\}-\text{log}_2|{\bf W}|\nonumber\\
&\quad\quad\text{s.t.}\quad\quad \text{(\ref{eq:problemb})}-\text{(\ref{eq:problemd})}.
\end{align}
\begin{IEEEproof}
See Appendix A.
\end{IEEEproof}
Note that while more optimization variables are introduced in problem $({\cal P}_1)$ compared to the original problem $(\cal P)$, i.e., ${\bf W}$, the reformulated problem $({\cal P}_1)$ is in a more tractable form. More specifically, for a given RIS reflection matrix $\bm\Theta$, $({\cal P}_1)$ is convex with regard to the other optimization variables. Consequently, we can solve the problem $({\cal P}_1)$ in a more effective way by using the alternating optimization approach.

To this end, we first derive the optimal solution for $\bf W$ when the other optimization variables are fixed. In that case, problem $({\cal P}_1)$ is reformulated as an unconstrained problem in terms of ${\bf W}$. Then by utilizing the KKT condition, the optimal $\bf W$ is give by
\begin{equation}\label{eq:W}
{\bf W}={\bf E}^{-1}.
\end{equation}
When $\bf W$ is fixed, the optimization problem $({\cal P}_1)$ becomes the weighted MMSE problem which is more tractable than the original rate maximization problem. For the remainder of
this section, we first pay attention to the optimization of ${\bf W}_s$ and ${\bf W}_d$ with fixed $\bm\Theta$, and then focus on the optimization of $\bm\Theta$ with fixed ${\bf W}_s$ and ${\bf W}_d$.
\subsection{Optimal Transceiver Design}
Next, we first deal with the optimization of the precoder ${\bf W}_s$ and then on the combiner ${\bf W}_d$ while fixing $\bm\Theta$. In particular, the corresponding problem with respect to ${\bf W}_s$ is written as
\begin{align}\label{eq:problem1}
&{\mathop{\text{min}}\limits_{_{{\bf W}_s}}}\quad\quad{\text{WMSE}_1}\nonumber\\
&~\text{s.t.}\quad\quad\text{(\ref{eq:problemb})},
\end{align}
where
\begin{align}
\text{WMSE}_1&\!\!=\!\!\text{tr}\{\bf WE\}\nonumber\\
&\!\!=\!\!{\text{tr}}\left\{\!{\sigma^2{\bf{WW}_d}^H{{\bf{W}_d}}} \!\right\}\!-\!2{\text {Re}} \left\{\!{{\text{tr}}\left\{ {{\bf{W}}{\bf{W}_s}^H{\bf{H}}_{\text{eq}}^H{{\bf{W}_d}}}\!\right\}} \right\}\nonumber\\
&~\!\!+{\text{tr}}\left\{ {{\bf{W}}} \right\}+ {\text{tr}}\left\{ {{\bf{WW}_d}^H{{\bf{H}}_{\text{eq}}}{{\bf{W}_s}}{\bf{W}_s}^H{\bf{H}}_{\text{eq}}^H{{\bf{W}_d}}} \right\}.\nonumber
\end{align}
Note that (\ref{eq:problem1}) is convex with regard to ${\bf W}_s$, we first derive its Lagrangian function:
\begin{equation}\label{eq:lagrangian}
\emph{L}({\bf W}_s,\mu) =\text{WMSE}+\mu\left(\text {tr}\left\{{\bf W}_s{\bf W}_s^H\right\}- P\right),
\end{equation}
with $\mu\geq 0$ being the Lagrangian multiplier associated with (\ref{eq:problem1}). We then differentiate (\ref{eq:lagrangian}) with respect to ${\bf W}_s$, which yields
\begin{equation}\label{eq:differentiate}
\frac{\partial L({\bf W}_s,\mu)}{\partial {\bf W}_s}={\bf H}_{\text{eq}}^H{\bf W}_d{\bf{WW}_d}^H{\bf H}_{\text{eq}}{\bf W}_s - {\bf H}_{\text{eq}}^H{\bf W}_d{\bf W} + \mu{\bf W}_s.
\end{equation}
Finally the optimal ${\bf W}_s$ is obtained by solving $\frac{\partial L({\bf W}_s,\mu)}{\partial {\bf W}_s} = {\bf 0}$, whose solution is
\begin{equation}\label{eq:Ws}
{\bf W}_s = \left({\bf H}_{\text{eq}}^H{\bf W}_d{\bf W}{\bf W}_d^H{\bf H}_{\text{eq}}  + {\mu}{\bf I }_{N_t} \right)^{ - 1}{\bf H}_{\text{eq}}^H{\bf W}_d{\bf W},
\end{equation}
and $\mu$ is chosen to satisfy the power constraint. Specifically, let ${\bf W}_s(\mu)$ denote the right-side hand of (\ref{eq:Ws}). When ${\bf H}_{\text{eq}}^H{\bf W}_d{\bf W}{\bf W}_d^H{\bf H}_{\text{eq}}$ is invertible and $\text {tr}\left({\bf W}_s(0){\bf W}_s(0)^H\right)\leq P$, then ${\bf W}_s={\bf W}_s(0)$, otherwise we must have
\begin{equation}\label{eq:power}
\text {tr}\left\{{\bf W}_s(\mu){\bf W}_s(\mu)^H\right\}= P,
\end{equation}
which can be written as
\begin{equation}\label{eq:mu1}
\text{tr}\{({\bm\Lambda}+\mu{\bf I}_{N_t})^{-2}{\bm\Phi}\}=P,
\end{equation}
where ${\bf U}{\Lambda }{\bf U}^H$ is the eigendecomposition of ${\bf H}_{\text{eq}}^H{\bf W}_d{\bf W}{\bf W}_d^H{\bf H}_{\text{eq}}$, and $\bm\Phi={\bf U}^H{\bf H}_{\text{eq}}^H{\bf W}_d{\bf W}^2{\bf W}_d^H{\bf H}_{\text{eq}}{\bf U}$. Therefore, (\ref{eq:mu1}) is rewritten as
\begin{equation}\label{eq:mu2}
\sum\limits_{i = 1}^{{N_t}}\frac{{\bm\Phi}_{i,i}}{({\bm\Lambda}_{i,i}+\mu)^2} =P,
\end{equation}
where ${\bm\Phi}_{i,i}$ and ${\bm\Lambda}_{i,i}$ are respectively the $i$th diagonal element of $\bm\Phi$ and $\bm\Lambda$.

Obviously, the left-side hand of  (\ref{eq:mu2}) is monotonically decreasing with respect to $\mu$. Recall that $\mu\geq 0$, hence, we can easily get the optimal $\mu$ by using bisection method. Finally, by substituting the optimal $\mu$ to (\ref{eq:Ws}), the closed-form solution to ${\bf W}_s$ is obtained directly.

Analogously, it can also be proven that the MMSE solution given in (\ref{eq:Wd}) is exactly the optimal solution to ${\bf W}_d$. Now that we have completed the optimization of ${\bf W}_s$ and ${\bf W}_d$, then we focus on problem $({\cal P}_1)$ for the optimization with respect to $\bm\Theta$.
\subsection{Design of the RIS Reflection Matrix}Now we turn again our attention on the unfinished task of problem $({\cal P}_1)$ with given ${\bf W}_s$ and ${\bf W}_d$. For the purpose of clarity in our presentation, we first define $\overline{\bf G}\triangleq{\bf W}_d^H{\bf G}$, ${\bf y}_R\triangleq{\bf H}{\bf W}_s{\bf s}$, $\overline{\bf H}_d\triangleq{\bf W}_d^H{{\bf{H}}_d}{{\bf{W}_s}}$, and $\overline{\bf n}\triangleq{\bf W}_d^H{\bf n}$. Accordingly, the signal model in (\ref{eq:combine}) is recast as
\begin{equation}\label{eq:y}
{{\bf y}} = \overline{\bf G}{\bm\Theta}{\bf y}_R + \overline{\bf H}_d{\bf{s}} + \overline{\bf n}.
\end{equation}
By fixing the other optimization variables and substituting (\ref{eq:y}) into the corresponding objective of $({\cal P}_1)$, the univariate optimization problem only with respect to $\bm\Theta$ is obtained as follows
\begin{align}\label{eq:problem2}
&{\mathop{\text{min}}\limits_{_{\bm\Theta}}}\quad\quad{\text{WMSE}_2}\nonumber\\
&~\text{s.t.}\quad\quad\text{(\ref{eq:problemc})},\text{(\ref{eq:problemd})},
\end{align}
with
\begin{align}
{\text{WMSE}_2}&=\mathbb{E}\left\{ {\left\| {{{\bf{W}}^{1/2}}\left( {{\bf{s}} - {{\overline {\bf{G}} }}{\bf{\Theta }}{{\bf{y}}_R} - {{\overline {\bf{H}} }_d}{\bf{s}} - \overline {\bf{n}} } \right)} \right\|_2^2}\right\}\nonumber\\
&=-2 \text{Re}\left\{\text{tr}\left\{\left({\bf I}-\overline{\bf H}_d\right){\bf R}_x{\bm\Theta}^H\overline{\bf G}^H{\bf W}\right\}\right\}\nonumber\\
&+\text{tr}\left\{\overline{\bf G}{\bm\Theta}{\bf R}_y{\bm\Theta}^H\overline{\bf G}^H{\bf W}\right\}\!+\!\text{tr}\left\{\overline{\bf H}_d\overline{\bf H}_d^H{\bf W}\right\}\nonumber\\
&~-2\text{Re}\left\{\text{tr}\left\{\overline{\bf H}_d^H{\bf W}\right\}\right\}+\text{tr}\left\{{\bf W}\right\}+\text{tr}\left\{\sigma^2{\bf W}_d^H{\bf W}_d{\bf W}\right\},\nonumber
\end{align}
where ${\bf R}_x\triangleq\mathbb{E}\left\{{\bf s}{\bf y}_R^H\right\}={\bf W}_s^H{\bf H}^H$ and ${\bf R}_y\triangleq\mathbb{E}\left\{{\bf y}_R{\bf y}_R^H\right\}={\bf H}{\bf W}_s{\bf W}_s^H{\bf H}^H$.

In order to facilitate the mathematical tractability, we further define ${\bf W}_x\triangleq\left({\bf I}-\overline{\bf H}_d\right){\bf R}_x{\bf R}_y^{-1}$, it yields
\begin{equation}\label{eq:wr}
\left({\bf I}-\overline{\bf H}_d\right){\bf R}_x{\bm\Theta}^H\overline{\bf G}^H{\bf W}={\bf W}_x{\bf R}_y{\bm\Theta}^H\overline{\bf G}^H{\bf W}.
\end{equation}
Then by introducing a constant term $\text{tr}\left\{{\bf W}_x{\bf R}_y{\bf W}_x^H{\bf W}\right\}$, we
can reformulate into WMSE$_2$ as follows
\begin{align}\label{eq:mse3}
{\text{WMSE}_2}&=-2 \text{Re}\left\{\text{tr}\left\{{\bf W}_x{\bf R}_y{\bm\Theta}^H\overline{\bf G}^H{\bf W}\right\}\right\}+\text{tr}\left\{{\bf W}\right\}\nonumber\\
&~+\text{tr}\left\{\overline{\bf G}{\bm\Theta}{\bf R}_y{\bm\Theta}^H\overline{\bf G}^H{\bf W}\right\}+\text{tr}\left\{{\bf W}_x{\bf R}_y{\bf W}_x^H{\bf W}\right\}\nonumber\\
&~+\text{tr}\left\{\overline{\bf H}_d\overline{\bf H}_d^H{\bf W}\right\}-\text{tr}\left\{{\bf W}_x{\bf R}_y{\bf W}_x^H{\bf W}\right\}\nonumber\\
&~-2\text{Re}\left\{\text{tr}\left\{\overline{\bf H}_d^H{\bf W}\right\}\right\}+\text{tr}\left(\sigma^2{\bf W}_d^H{\bf W}_d{\bf W}\right)\nonumber\\
&=\left\| {\bf W}^{1/2}{\bf W}_x{\bf R}_y^{1/2}-{\bf W}^{1/2}\overline{\bf G}{\bm\Theta}{\bf R}_y^{1/2}\right\|_{\text F}^2+c,
\end{align}
where $c$ is a constant that includes the extraneous terms from the first equation. After securely omitting this constant term, the WMMSE-based RIS design for (\ref{eq:problem2}) with given ${\bf W}_s$ and ${\bf W}_d$ can be equivalently obtained by solving
\begin{align}\label{eq:problem3}
&{\mathop{\text{min}}\limits_{_{{\bm\Theta}}}}\quad {\left\| {\bf W}^{1/2}{\bf W}_x{\bf R}_y^{1/2}-{\bf W}^{1/2}\overline{\bf G}^H{\bm\Theta}{\bf R}_y^{1/2}\right\|_{\text F}^2 }\nonumber\\
&~{\text{s.t.}}\quad~\text{(\ref{eq:problemc})},\text{(\ref{eq:problemd})},
\end{align}
where
\begin{align*}
\begin{split}
\left\{ \begin{array}{l}
{\bf W}_x\triangleq\left({\bf I}-\overline{\bf H}_d\right)\mathbb{E}\left\{{\bf s}{\bf y}_R^H\right\}\mathbb{E}\left\{{\bf y}_R{\bf y}_R^H\right\}^{-1}\\
\quad\quad=\left({\bf I}-\overline{\bf H}_d\right)\overline{\bf H}^H\left(\overline{\bf H}\overline{\bf H}^H\right)^{-1},\\
{\bf R}_y\triangleq\mathbb{E}\left\{{\bf y}_R{\bf y}_R^H\right\}=\overline{\bf H}\overline{\bf H}^H,
\end{array} \right.
\end{split}
\end{align*}
with $\overline{\bf H}\triangleq{\bf H}{\bf W}_s$.

To proceed further, we then convert the matrix operation in (\ref{eq:problem3}) to the equivalent vector operation and eliminate the diagonal constraint (\ref{eq:problemc}) by utilizing the property of Kronecker product. Accordingly, (\ref{eq:problem3}) can be rewritten as
\begin{align}\label{eq:problem4}
&{\mathop{\text{min}}\limits_{_{{\bm\Theta}}}}\quad \left\| {\bf a}_r\!-{\bf A}\text{vec}\left({\bm\Theta}\right)\right\|_{2}^2\nonumber\\
&~{\text{s.t.}}\quad~\text{(\ref{eq:problemc})},\text{(\ref{eq:problemd})},
\end{align}
where
\begin{align*}
\begin{split}
\left\{ \begin{array}{l}
{{\bf{a}}_r} \buildrel \Delta \over = {\rm{vec}}\left( {{{\bf{W}}^{1/2}}{{\bf{W}}_x}{\bf{R}}_y^{1/2}} \right),\\
{\bf{A}} \buildrel \Delta \over = {\bf{R}}_y^{1/2} \otimes \left( {{{\bf{W}}^{1/2}}{{\overline {\bf{G}} }^H}} \right).
\end{array} \right.
\end{split}
\end{align*}
Next, we extract the non-zero elements of $\text{vec}\left({\bm\Theta}\right)$ to remove the diagonal constraint in (\ref{eq:problem4}) according to the following lemma.

\emph{Lemma 2:} For ${\bf N}=\left[{\bf n}_1,{\bf n}_2,\cdots,{\bf n}_N\right]\in{\mathbb C}^{M\times N}$ and ${\bf r}=\left[r_1,r_2,\cdots,r_N\right]^T\in{\mathbb C}^{N\times 1}$. If $r_i=0$, then we have
\begin{equation}\label{eq:lemma1}
{\bf N}{\bf r} = \hat{\bf N}\hat{\bf r},
\end{equation}
in which $\hat{\bf N}=\left[{\bf n}_1,\cdots,{\bf n}_{i-1},{\bf n}_{i+1},\cdots,{\bf n}_N\right]\in{\mathbb C}^{M\times(N-1)}$ and $\hat{\bf r}=\left[r_1,\cdots,r_{i-1},r_{i+1},\cdots,r_N\right]^T\in{\mathbb C}^{(N-1)\times 1}$.
\begin{IEEEproof}
This can be easily proven according to the basic matrix multiplications.
\end{IEEEproof}
By applying \emph{Lemma 2}, we obtain the following equation
\begin{equation}\label{eq:change}
{\bf A}\text{vec}\left({\bm\Theta}\right)={\bf A}_r{\bm\theta},
\end{equation}
in which ${\bf A}_r$ can be produced by eliminating the associated columns from ${\bf A}$, i.e., ${\bf A}_r=\left[{\bf a}_1,{\bf a}_{N+2}, \cdots,{\bf a}_{N^2}\right]$, and ${\bf a}_i$ denotes the $i$th column of ${\bf A}$. Mathetically, ${\bf A}_r$ can be defined as
\begin{equation}\label{eq:Ar}
{{\bf{A}}_r} = \left[ {{{\bf{r}}_1} \otimes {{\bf{g}}_1},{{\bf{r}}_2} \otimes {{\bf{g}}_2}, \cdots ,{{\bf{r}}_N} \otimes {{\bf{g}}_N}} \right],
\end{equation}
where ${\bf{r}}_n$ and ${\bf{g}}_n$ denote the $n$th column of ${\bf{R}}_y^{1/2}$ and ${{\bf{W}}^{1/2}}{\overline {\bf{G}} ^H}$, respectively, $\forall n \in \left\{ {1,2, \cdots ,N} \right\}$.

Plugging (\ref{eq:change}) into (\ref{eq:problem4}), it is reduced to
\begin{align}\label{eq:problem5}
&{\mathop{\text{min}}\limits_{_{{\bm\theta}}}}\quad \left\| {\bf a}_r\!-{\bf A}_r{\bm\theta}\right\|_{2}^2\nonumber\\
&~\text{s.t.}\quad~\text{(\ref{eq:problemd})}.
\end{align}
Mathematically, by elaborating the objective function of problem (\ref{eq:problem5}), (\ref{eq:problem5}) is equivalent to:
\begin{align}\label{eq:problem6}
&{\mathop{\text{min}}\limits_{_{{\bm\theta}}}}\quad {\bf a}_r^H{\bf a}_r+{\bm\theta}^H{\bf A}_r^H{\bf A}_r{\bm\theta}-2\text{Re}\left\{{\bm\theta}^H{\bf A}_r^H{\bf a}_r\right\}\nonumber\\
&~\text{s.t.}\quad\text{(\ref{eq:problemd})}.
\end{align}
Up to now, while the alternative problem formulation in (\ref{eq:problem6}) is yet non-convex, it is easier to be handled by existing optimization tools after a series of mathematical transformations. For the remainder of this subsection, two computationally efficient methods are advocated to tackle (\ref{eq:problem6}).
\subsubsection{Semi-Definite Relaxation} Problem (\ref{eq:problem6}) is a standard QCQP with unit-modulus constraints, which can be reformulated as
\begin{align}\label{eq:problem7}
&{\mathop{\text{min}}\limits_{_{{\overline {\bm\theta}}}}}\quad \overline{\bm\theta}^H{\bf R}_r\overline{\bm\theta}\nonumber\\
&\text{s.t.}\quad|\overline\theta_n|=1,\forall n\in\{1,2,\cdots,N+1\},
\end{align}
where
\begin{equation}
{\bf R}_r = \left[ {\begin{array}{*{20}{c}}
{\bf A}_r^H{\bf A}_r&{\bf A}_r^H{\bf a}_r\\
{\bf a}_r^H{\bf A}_r&{\bf a}_r^H{\bf a}_r
\end{array}} \right],\overline{\bm\theta} = \left[ {\begin{array}{*{20}{c}}
{\bm\theta}\\
{1}
\end{array}} \right].
\end{equation}
Define $\overline{\bm\Theta}\triangleq\overline{\bm\theta}\overline{\bm\theta}^H$, problem (\ref{eq:problem7}) is equivalent to
\begin{align}\label{eq:problem8}
&{\mathop{\text{min}}\limits_{_{{\overline {\bm\Theta}}\succeq{\bf 0}}}}\quad \text{tr}\{\overline{\bm\Theta}{\bf R}_r\}\nonumber\\
&\text{s.t.}\quad\overline{\bm\Theta}_{n,n}=1,\forall n\in\{1,2,\cdots,N+1\}\nonumber\\
&\quad\quad\text{rank}(\overline{\bm\Theta})=1.
\end{align}
Considering that the above rank-one constraint is non-convex, we temporarily relax it by utilizing SDR technique, which yields
\begin{align}\label{eq:problem9}
&{\mathop{\text{min}}\limits_{_{{\overline {\bm\Theta}}}}}\quad \text{tr}\{\overline{\bm\Theta}{\bf R}_r\}\nonumber\\
&\text{s.t.}\quad\overline{\bm\Theta}_{n,n}=1,\forall n\in\{1,2,\cdots,N+1\}\nonumber\\
&\quad~\quad{\overline {\bm\Theta}}\succeq{\bf 0}.
\end{align}
Obviously, (\ref{eq:problem9}) is a semi-definite programming (SDP) problem and hence can be tackled effortlessly by CVX. Moreover, to recover the rank-one $\bm\theta$ from the higher-rank $\bm\Theta$, Gaussian randomization is recommended, the details of which have been investigated in \cite{wu2019towards} and hence ignored here. However, it is noted that if the Gaussian randomization trials are not sufficient, the optimality of $\bm\Theta$ would no longer exist and thus the proposed algorithm would not converge. Hence, to guarantee its convergence, a large number of random trials are required, causing extremely high computational complexity. To address this issue, we then advocate another computationally-efficient method.
\subsubsection{Sequence of Closed Forms}The SCF method solves a difficult non-convex optimization problem with constant-modulus constraint by tackling a series of convex equality constrained QP problems instead. Specifically, each of the subproblem turns out to have a closed form solution and the constant-modulus is satisfied upon convergence \cite{aldayel2017tractable}.

For complex-valued problem (\ref{eq:problem6}), its equivalent real-valued problem is formulated as follows:
\begin{align}\label{eq:problem7}
&{\mathop{\text{min}}\limits_{_{{\bf u}}}}\quad {\bf u}^T{\bf P}{\bf u}-{\bf t}^T{\bf u}-{\bf u}^T{\bf t}+ r\nonumber\\
&~\text{s.t.}\quad u_n^2+u_{n+N}^2=1,~\forall n\in\{1,2, \cdots ,N\}
\end{align}
where $r\triangleq{\bf a}_r^H{\bf a}_r,~{\bf u}\triangleq\left[{\text{Re}}\{\bm\theta\}^T,{\text{Im}}\{\bm\theta\}^T\right]^T$, $u_l$ is the $l$th element of $\bf u$, and
\begin{equation}\nonumber
{\bf{P}}\!\triangleq\!\left[\!{\begin{array}{*{20}{l}}\!
{{\text{Re}}\{ {\bf{A}}_r^H{{\bf{A}}_r}\} }\!&\!{ - {\text{Im}}\{ {\bf{A}}_r^H{{\bf{A}}_r}\} }\\
\!{{\text{Im}}\{ {\bf{A}}_r^H{{\bf{A}}_r}\}}\!&\!{{\text{Re}}\{ {\bf{A}}_r^H{{\bf{A}}_r}\} }
\end{array}}\!\!\right],{\bf{t}}\!\triangleq\!\left[ {\begin{array}{*{20}{l}}\!
{{\text{Re}}\{ {\bf{A}}_r^H{{\bf{a}}_r}\}\!\!}\\
\!{{\text{Im}}\{ {\bf{A}}_r^H{{\bf{a}}_r}\}\!\!}
\end{array}} \right].
\end{equation}
Define
\begin{equation}\nonumber
{\bf{R}}\triangleq\left[{\begin{array}{*{20}{l}}
~{\bf P}&{ - \bf t }\\
\!{-\bf t}^T&~{r}
\end{array}}\!\!\right],{\bf{x}}\triangleq\left[ {\begin{array}{*{20}{l}}
{\bf u}\\
{1}
\end{array}} \right]=\left[ {\begin{array}{*{20}{l}}
{{\text{Re}}\{ \bm\theta\}}\\
{{\text{Im}}\{ \bm\theta\}}\\
\quad 1
\end{array}} \right],
\end{equation}
then the objective of (\ref{eq:problem7})is reexpressed as
\begin{equation}\label{eq:convert}
{\bf u}^T{\bf P}{\bf u}-{\bf t}^T{\bf u}-{\bf u}^T{\bf t}+ r={\bf x}^T{\bf R}{\bf x}.
\end{equation}
By substituting (\ref{eq:convert}), (\ref{eq:problem7}) can be equivalently written as
\begin{align}\label{eq:problem8}
&{\mathop{\text{min}}\limits_{_{{\bf x}}}}\quad {{{\bf{x}}^T}({\bf{R}} + \lambda {\bf{I}}_{2N+1}){\bf{x}}}\nonumber\\
&~\text{s.t.}\quad {{{\bf{x}}^T}{{\bf{E}}_n}{\bf{x}} = 1},~\forall n\in\{ 1,2, \cdots ,N+1\}
\end{align}
where $\lambda>0$ is a auxiliary variable and ${\bf{E}}_n$ is a $(2N+1)\times (2N+1)$ matrix which is defined as
\begin{equation}\nonumber
{{\bf{E}}_n}(m,l) = \left\{ {\begin{array}{*{20}{l}}
1&{{\text{ if }}~ m = l = n,n \le N},\\
1&{{\text{ if }}~ m  =n,l= n + N,n \le N},\\
1&{{\text{ if }}~ m = l = 2N + 1,n = N + 1},\\
0&{{\text{ Otherwise}}.}
\end{array}} \right.
\end{equation}
Note that since (\ref{eq:problem8}) enforces ${{{\bf{x}}^T}{{\bf{E}}_n}{\bf{x}} = 1},~\forall n\in\{ 1,2, \cdots ,N+1\}$, then we have $\lambda{\bf x}^T{\bf x}=\lambda(N+1)$. Consequently, the optimal solution to (\ref{eq:problem8}) and that to (\ref{eq:problem6}) (the complex version of (\ref{eq:problem7})) are completely identical for any $\lambda\geq 0$.

Next, we focus on the relaxation of the unit-modulus constraint of (\ref{eq:problem8}). Consider the following sequences of constraints:
\begin{equation}\label{eq:constraint}
{\bf B}^{(i)}{\bf x}={\bf 1},
\end{equation}
with
\begin{small}
\begin{equation}\label{eq:Bn}
{{\bf{B}}^{(i)}}(m,l)\!=\!\left\{ {\begin{array}{*{20}{l}}
\!\!{\cos \left( {\arg \left( {\theta _n^{(i - 1)}} \right)}\right)} &\!\!{{\text{ if }}~ m = l = n,n \le N},\\
\!\!{\sin \left( {\arg \left( {\theta _n^{(i - 1)}} \right)} \right)}&\!\!{{\text{ if }}~ m = n,l = n + N,n \le N,}\\
\!\!1&\!\!{{\text{ if }}~ m = N + n,l = 2N + 1,}\\
\!\!0&\!\!{{\text{ Otherwise}}.}
\end{array}} \right.
\end{equation}
\end{small}where $\theta_n^{(i - 1)}=x_n^{(i - 1)}+jx_{n+N}^{(i - 1)}$ with $x_n^{(i - 1)}$ being the $n$th element of ${\bf x}$ obtained in the $(i-1)$th iteration.

Replacing the unit-modulus constraint in (\ref{eq:problem8}) by (\ref{eq:constraint}), we obtain the subsequent sequence of QPs subject to equality constraints
\begin{align}\label{eq:qp}
({\cal {QP}}^{(i)})\quad&{\mathop{\text{min}}\limits_{_{{\bf x}}}}\quad {{{\bf{x}}^T}({\bf{R}} + \lambda {\bf{I}}_{2N+1}){\bf{x}}}\nonumber\\
&~\text{s.t.}\quad {\bf B}^{(i)}{\bf x}={\bf 1},
\end{align}
whose optimal solution is easily acquired by its optimality condition according to \cite{boyd2004convex}:
\begin{align}\label{eq:solution}
{\bf x}^{(i)}=\overline{\bf R}^{-1}{{\bf B}^{(i)}}^T\left({\bf B}^{(i)}\overline{\bf R}^{-1}{{\bf B}^{(i)}}^T\right)^{-1}{\bf 1},
\end{align}
where $\overline{\bf R}=2({\bf R}+\lambda{\bf I}_{2N+1})$.

Specifically, while the problem $({\cal {QP}}^{(i)})$ does not lead to a unit-modulus solution, a series of $({\cal {QP}}^{(i)})$ generates a sequence of non-increasing objective values. Furthermore, the converged solution is guaranteed to satisfy the unit modulus. In other words, the affine constraint (\ref{eq:constraint}) is adjusted to satisfy the unit-modulus constraint. To explain this, let ${\bm\theta}^{(i)}$ be the complex version of ${\bf x}^{(i)}$, i.e., ${\bf x}^{(i)}=\left[{\text{Re}}\{{\bm\theta}^{(i)}\}^T,{\text{Im}}\{{\bm\theta}^{(i)}\}^T,1\right]^T$. If ${\bm\theta}^{(i)}$ satisfies the unit-modulus constraint, then we have ${\bm\theta}^{(i)}={\bm\theta}_{(i-1)}$, where ${\bm\theta}_{(i-1)}$ denotes the unit-modulus version of ${\bm\theta}^{(i-1)}$ with ${\bm\theta}_{(i-1)}=e^{\jmath\text{arg}({\bm\theta}^{(i-1)})}$. Then we get ${\bf B}^{(i+1)}={\bf B}^{(i)}$, which indicates that the constraint of problem $({\cal {QP}}^{(i+1)})$ is the same as that of $({\cal {QP}}^{(i)})$. Consequently, ${\bm\theta}^{(i+1)}={\bm\theta}^{(i)}$ is obtained, the iteration converges. Otherwise, the constraint will be continuously updated according to (\ref{eq:constraint}). Together, the convergence is then guaranteed by the following lemma.

\emph{Lemma 3:} The objective function value of the sequence of (\ref{eq:problem6}) with respect to ${\bm\theta}_{(i)}$ is non-increasing in $i$ when $\lambda\geq\frac{N}{8}\lambda_{\text{max}}({\bf A}_r^H{\bf A}_r)+\|{\bf A}_r^H{\bf a}_r\|_2$. Moreover, upon convergence, unit modulus is guaranteed and the unit-modulus solution satisfies the KKT conditions of problem (\ref{eq:problem6}).
\begin{IEEEproof}
See Appendix B.
\end{IEEEproof}
\subsection{Computational Complexity}
Up to now we have finished the optimizations of ${\bf W},{\bf W}_s,{\bf W}_d$, and $\bm\Theta$ in Section \uppercase\expandafter{\romannumeral3}, our two proposed
achievable rate maximization algorithms for the considered system are respectively summarized in Algorithms 1
and 2, where we initialize ${\bf W}_s$ by the singular value decomposition of the effective channel, i.e., ${\bf H}_{\text{eq}}$. As shown, the optimization variables are iteratively and alternatively obtained until convergence. On the one hand, both the SDR-based and the SCF-based approaches increase the capacity after each iteration. One the other hand, the achievable rate has been upper-bounded within the feasible set of (\ref{eq:problemb})-(\ref{eq:problemd}), as a result it will not grow infinitely. Together, the convergence of both Algorithm 1 and Algorithm 2 are guaranteed.

The associated complexity of the two proposed algorithms mainly
hinges on two parameters, i.e., the number of external iterations, say $T$, and the complexity required to obtain each variable.

As for Algorithm 1, the optimization of $\bm\Theta$ is based on SDR with Gaussian randomization, whose computational complexity is of ${\cal O}(N^{3.5})+{\cal O}(LN^2)$, with $L$ being the number of randomization trials. To sum up, the complexity of Algorithm 1 can be evaluated as
\begin{equation}\label{eq:complexity1}
{\cal O}\left(T(N_t^3+N^{3.5}+LN^2)\right).
\end{equation}

A similar analysis applies to Algorithm 2, except that the computational complexity of requiring the optimal RIS reflection matrix is ${\cal O}(IN^{2.373})$ where $I$ is the number of SCF iterations. Analogously, the complexity of Algorithm 2 is calculated as
\begin{equation}\label{eq:complexity2}
{\cal O}\left(T(N_t^3+IN^{2.373}+N^2N_r)\right).
\end{equation}
Generally, the number of Gaussian randomization trials $L$ is far greater than the number of RIS reflecting elements and the number of SCF iterations, i.e., $L>>N,L>>I$, thus, the SCF-based method, or Algorithm 2 exhibits significantly reduced complexity in contrast to Algorithm 1.
\begin{algorithm}[t]
\caption{SDR-based Capacity Maximization Algorithm}
\label{alg:Framwork}
\begin{algorithmic}[1]
\STATE Let $t=0$ and initialize ${\bm\theta}^{(t)}$, ${\bf W}_s^{(t)}$. Let ${\bm\Theta}^{(t)}=\text{diag}\left({\bm\theta}^{(t)}\right),{\bf H}_{\text{eq}}^{(t)}\triangleq {\bf G}^H{\bm\Theta}^{(t)}{\bf H}+{\bf H}_d$.
\REPEAT
\STATE$t \leftarrow  t+1$
\STATE Calculate ${\bf W}_d^{(t)}$ according to (\ref{eq:Wd});
\STATE Calculate ${\bf W}^{(t)}$ according to (\ref{eq:W});
\STATE Calculate ${\bf W}_s^{(t)}$ according to (\ref{eq:Ws});
\STATE Solve (\ref{eq:problem9}) using CVX optimization tool box, and apply Gaussian randomization to obtain ${\bm\theta}^{(t)}$.
\STATE ${\bm\Theta}^{(t)}=\text{diag}\left({\bm\theta}^{(t)}\right)$;
\STATE ${\bf H}_{\text{eq}}^{(t)}\triangleq {\bf G}^H{\bm\Theta}^{(t)}{\bf H}+{\bf H}_d$;
\UNTIL{convergence}
\STATE ${\bf W}={\bf W}^{(t)},{\bf W}_s={\bf W}_s^{(t)},{\bf W}_d={\bf W}_d^{(t)},{\bm\Theta}={\bm\Theta}^{(t)}$.
%\RETURN ${\bm\theta},{\bf G}$; %算法的返回值
\end{algorithmic}
\end{algorithm}
\begin{algorithm}[t]
\caption{SCF-based Capacity Maximization Algorithm}
\label{alg:Framwork}
\begin{algorithmic}[1] %这个1 表示每一行都显示数字
%\REQUIRE ~~\\ %算法的输入参数：Input
\STATE Let $t=0$ and initialize ${\bm\theta}^{(t)}$, ${\bf W}_s^{(t)}$. Let ${\bm\Theta}^{(t)}=\text{diag}\left({\bm\theta}^{(t)}\right),{\bf H}_{\text{eq}}^{(t)}\triangleq {\bf G}^H{\bm\Theta}^{(t)}{\bf H}+{\bf H}_d$.
\REPEAT
\STATE$t \leftarrow  t+1$
\STATE Calculate ${\bf W}_d^{(t)}$ according to (\ref{eq:Wd});
\STATE Calculate ${\bf W}^{(t)}$ according to (\ref{eq:W});
\STATE Calculate ${\bf W}_s^{(t)}$ according to (\ref{eq:Ws});
\STATE $i=0,{\bm\theta}^{i}={\bm\theta}^{(t-1)},{\bm\theta}_i=e^{\jmath\text{arg}({\bm\theta}^{i})}$;
\WHILE{($|h({\bm\theta}_i)-h({\bm\theta}_{i-1})|\geq\epsilon$}
\STATE $i \leftarrow  i+1$;
\STATE Calculate ${\bf B}^{(i)}$ according to (\ref{eq:Bn});
\STATE ${\bf x}^{(i)}=\overline{\bf R}^{-1}{{\bf B}^{(i)}}^T\left({\bf B}^{(i)}\overline{\bf R}^{-1}{{\bf B}^{(i)}}^T\right)^{-1}{\bf 1}$;
\STATE ${\bm\theta}^i=\left[{\bf x}^{(i)}\right]_{(1:N)}+\jmath\left[{\bf x}^{(i)}\right]_{(N+1:2N)}$;
\STATE ${\bm\theta}_i=e^{\jmath\text{arg}\{{\bm\theta}^{i}\}}$;
\ENDWHILE
\STATE ${\bm\theta}^{(t)}={\bm\theta}^i$;
\STATE ${\bm\Theta}^{(t)}=\text{diag}\left({\bm\theta}^{(t)}\right)$;
\STATE ${\bf H}_{\text{eq}}^{(t)}\triangleq {\bf G}^H{\bm\Theta}^{(t)}{\bf H}+{\bf H}_d$.
\UNTIL{convergence}
\STATE ${\bf W}={\bf W}^{(t)},{\bf W}_s={\bf W}_s^{(t)},{\bf W}_d={\bf W}_d^{(t)},{\bm\Theta}={\bm\Theta}^{(t)}$.
\end{algorithmic}
\end{algorithm}
\subsection{Convergence and Optimality Analysis}
Denote the objective function of $({\cal{P}}_1)$ by $f\left({\bf W}_d,{\bf W},{\bf W}_s,{\bm\Theta}\right)$. First, it is obvious that the global optimality of ${\bf W}_d,{\bf W}$, and ${\bf W}_s$ can be guaranteed according to (\ref{eq:Wd}), (\ref{eq:W}) and (\ref{eq:Ws}). Besides, the local optimality of $\bm\Theta$ in Algorithm 2 has been proven in \emph{Lemma 2}. According to \cite{so2007approximating}, with a sufficiently large number of randomization trials, $\frac{\pi}{4}$-approximation of the optimal $\bm\Theta$ is guaranteed for SDR-based Algorithm 1. As a result, both Algorithm 1 and Algorithm 2 monotonically decrease the WMSE value after each iteration, which gives
\begin{small}
\begin{align}\label{eq:iteration}
&f\left({\bf W}_d^{(t-1)},{\bf W}^{(t-1)},{\bf W}_s^{(t-1)},{\bm\Theta}^{(t-1)}\right)\nonumber\\
&\geq f\left({\bf W}_d^{(t)},{\bf W}^{(t-1)},{\bf W}_s^{(t-1)},{\bm\Theta}^{(t-1)}\right)\nonumber\\
&\geq f\left({\bf W}_d^{(t)},{\bf W}^{(t)},{\bf W}_s^{(t-1)},{\bm\Theta}^{(t-1)}\right)\nonumber\\
&\geq f\left({\bf W}_d^{(t)},{\bf W}^{(t)},{\bf W}_s^{(t)},{\bm\Theta}^{(t-1)}\right)\!\geq \! f\left({\bf W}_d^{(t)},{\bf W}^{(t)},{\bf W}_s^{(t)},{\bm\Theta}^{(t)}\right).
\end{align}
\end{small}

Furthermore, $f\left({\bf W}_d^{(t)},{\bf W}^{(t)},{\bf W}_s^{(t)},{\bm\Theta}^{(t)}\right)$ can not decrease indefinitely since it is lower-bounded by zero. Hence, Algorithm 1 and Algorithm 2 always converge.

On the other hand, considering the optimality analysis, due to the existence of Gaussian randomization, no optimality claim can be made to Algorithm 1. So we only focus on the optimality analysis of Algorithm 2 based on the convergence analysis above.

To begin with, we rewrite problem $({\cal{P}}_1)$ as the following equivalent form
\begin{subequations}
\begin{align}
({\cal P}_2)~&{\mathop{\text{min}}\limits_{_{{\bf W}_s,{\bf W}_d,{\bf W},{\bf x}}}}\text{tr}\{{\bf WE}\}-\text{log}_2|{\bf W}|+\lambda{\bf x}{\bf x}^T\label{eq:p2a}\\
&~\quad\text{s.t.}\quad~\text{(\ref{eq:problemb})},\label{eq:p2b}\\
&\quad\quad\quad\quad{{{\bf{x}}^T}{{\bf{E}}_n}{\bf{x}} = 1},~\forall n\in\{ 1,2, \cdots ,N+1\}\label{eq:p2c}\\
&\quad\quad\quad\quad {\bm\Theta}=\text{diag}\left({\bf x}+\jmath{\bf x}\right)\label{eq:p2d}.
\end{align}
\end{subequations}
Note that problem $({\cal{P}}_2)$ is absolutely equivalent to problem $({\cal{P}}_1)$ since we only convert the complex-valued optimization variable $\bm\theta$ into its real-valued version $\bf x$, and add a constant term $\lambda{\bf x}{\bf x}^T=\lambda(N+1)$. Moreover, building upon the derivations in Section \uppercase\expandafter{\romannumeral3}-A, it is known that problem $(\cal{P})$ and problem $({\cal{P}}_1)$ are equivalent in the sense that the global optimal solutions of ${\bf W}_s$ and $\bm\Theta$ for the two problems hold the same. Hence, problem $({\cal{P}}_2)$ is also equivalent to $(\cal{P})$. As such, we present the optimality analysis of Algorithm 2 by analyzing the first-order necessary conditions of problem $({\cal{P}}_2)$.

\emph{Proposition 1:} As the increasing of the iteration $t$, Algorithm 2 is guaranteed to converge to a KKT point of problem $({\cal{P}}_1)$. In other words, the proposed Algorithm 2 is KKT optimal.
\begin{IEEEproof}
Since the diagonal constraint in (\ref{eq:p2d}) is naturally satisfied at the reformulation of the problem by utilizing \emph{Lemma 1}, we just consider the transmit power constraint (\ref{eq:p2b}) and the unit-modulus constraint (\ref{eq:p2c}). The KKT conditions of problem $({\cal{P}}_2)$ are expressed as follows
\begin{align}\label{eq:KKT}
&{\bf H}_{\text{eq}}^H{\bf W}_d{\bf{WW}_d}^H{\bf H}_{\text{eq}}{\bf W}_s - {\bf H}_{\text{eq}}^H{\bf W}_d{\bf W} + \mu{\bf W}_s={\bf 0},\nonumber \\
&2({\bf R}+\lambda {\bf I}_{2N+1}){\bf x}+\sum_{n=1}^{N+1} 2 \eta_n {\bf E}_n {\bf x} ={\bf 0}, \nonumber\\
&{{\bf x}}^{T}{\bf E}_n{\bf x}=1. \quad n \in\{1,2, \cdots, N+1\}
\end{align}
Suppose that the sequence of variables $\left\{{\bf W}_d^{(t)},{\bf W}^{(t)},{\bf W}_s^{(t)},{\bm\Theta}^{(t)}\right\}$ converges at the $t$th iteration. Evidently, the first equation can be guaranteed for ${\bf W}^{(t)}$ with the global optimality of ${\bf W}_s$. Besides, by following similar derivations as (\ref{eq:KKT1})-(\ref{eq:KKT3}) in the Appendix, the last equations can also be verified for ${\bf x}^{(t)}$. Recall that problem $({\cal{P}}_2)$ is equivalent to problem $({\cal{P}}_1)$, it can then be concluded that the final converged solution generated by Algorithm 2 meets the first-order optimality conditions of problem $({\cal{P}}_1)$.
\end{IEEEproof}
\section{Simulation Results}
In this section, numerical results are demonstrated to show the efficiency of our proposed algorithms, which solved the joint transceiver and reflecting beamforming design for the point-to-point MIMO system assisted by RIS. The error tolerance factor are chosen as $10^{-4}$ to guarantee the convergence rate of the proposed algorithms.

A uniform rectangular array (URA) of size $N=N_xN_y$ is assumed for the RIS, where $N_x$ and $N_y$ stand for the number of reflecting elements in the horizontal and vertical directions, respectively. For both the BS and the UE, we consider a uniform linear array (ULA). For all the channels involved, including ${\bf H},~ {\bf G}$, and ${\bf H}_d$, the  mmWave model is employed. For the purpose of precisely modeling the high-frequency characteristics, the well-known Saleh-Valenzuela model is adopted \cite{el2014spatially,zhao2022cooperative}. Take the channel from BS to the RIS, i.e., ${\bf H}$, as an example, it is given by
\begin{equation}\label{eq:los}
{{\bf{H}}} = \sqrt{\frac{N_tN}{N_{cl}N_{p}}}\sum\limits_{i\ell } {{\alpha _{il }^{\rm I}}}{{\bf{a}}_{\rm{I}}}\left( {\phi _{il}^{\rm{a}},\phi _{il}^{\rm{e}}} \right){{\bf{a}}_{\rm{t}}}{\left( {\phi _{il }^{\rm{t}}} \right)^H},
\end{equation}
where $N_{cl}$ denotes the total amount of scatters, and each of which has $N_p$ propagation paths. Here for simplicity we assume that each individual channel has the same number of scatters and paths, i.e., $N_{cl}=8$ and $N_{p}=10$. $\alpha_{il}\sim{\cal {CN}}(0,1)$ is the complex Gaussian gain corresponds to the $(i,l)$th link, i.e., the $l$th path in the $i$th scatter. Besides, $\phi_{il}^{\rm a}$ and $\phi_{il}^{\rm e}$ are respectively the azimuth and elevation angles of arrival (AOA) at the RIS, and $\phi_{il}^t$ denotes the angle of departure (AOD) at the BS. As for ${\bf a}_r(\phi_{il}^{\rm a} ,\phi_{il}^{\rm e})$ and ${\bf a}_t(\phi_{il}^t)$, they represent the corresponding array response vector for the URA and the ULA, respectively. Specifically, the expression of ULA is given by
\begin{equation}\label{eq:ula}
{{\bf{a}}_t}({\phi _t})\!=\!\frac{1}{{\sqrt {{N_t}} }}{\left[ {1,{{\text{e}}^{ - j\frac{{2\pi d}}{L }\sin \left( {{\phi _t}} \right)}}, \ldots ,{{\text{e}}^{ - j({N_t} - 1)\frac{{2\pi d}}{L }\sin \left( {{\phi _t}} \right)}}} \right]^T}\!,
\end{equation}
in which $L$ refers to the signal wavelength, and $d$ is the normalized antenna spacing. As for the UPA, the array response vector ${\bf a}_{\text I}(\phi _{\text I}^a,\phi_{\text I}^e)$ is given by [18]
\begin{equation}\label{eq:upa}
{\bf a}_{\text I}(\phi _{\text I}^a,\phi_{\text I}^e)= {\bf a}_{\text I}^a(\phi _{\text I}^a)\otimes{\bf a}_{\text I}^e(\phi _{\text I}^e),
\end{equation}
with ${\bf a}_{\text I}^a(\phi _{\text I}^a)\in{\mathbb C}^{N_x\times 1}$ and ${\bf a}_{\text I}^e(\phi _{\text I}^e)\in{\mathbb C}^{N_y\times 1}$ defined in the same manner as ${{\bf{a}}_{\text{t}}}({\phi _{\text{t}}})$.

As for the direct channel from the BS to the UE, it can be written as
\begin{equation}\label{eq:hd}
{{\bf{H}}_d} = \sqrt {\frac{{{N_t}{N_r}}}{{{N_{cl}}{N_{p}}}}} \sum\limits_{i = 1}^{{N_{cl}}} {\sum\limits_{l = 1}^{{N_{p}}} {{\alpha_{il}^{\rm d}}} } {{\bf{a}}_{\text{r}}}\left( {{\vartheta _{il}^{\rm r}}} \right){{\bf{a}}_{\text{t}}}{\left( {{\upsilon_{il}^{\rm t}}} \right)^H},
\end{equation}
where $\alpha_{il}^{\rm d}\sim {\cal {CN}}(0,1)$ is the corresponding path gain of the $(i,l)$th link. ${\bf{a}}_{\text r}\left( \vartheta _{il}^{\rm r}\right)$ and ${\bf{a}}_{\text t}\left(\varphi _{il}^{\rm t}\right)$ represent the receive and transmit array response vectors corresponding to the $(i,l)$th path, respectively, with $\vartheta _{il}^{\rm r}(\varphi _{il}^{\rm t})$ being the AoA/AoD. The array response vector corresponds to the ULA at the UE is in the same way as that modeled in (\ref{eq:ula}). Analogously, we denote the UE-RIS channel by
\begin{equation}\label{eq:ue}
{\bf G}= \sqrt {\frac{{{N}{N_r}}}{{{N_{cl}}{N_{p}}}}} \sum\limits_{i = 1}^{{N_{cl}}} {\sum\limits_{l = 1}^{{N_{p}}} {{\beta_{il}}} } {\bf a}_{\text I}(\psi _{il}^a,\psi_{il}^e){{\bf{a}}_{\text{U}}}{\left( {{\upsilon_{il}^{\rm r}}} \right)^H},
\end{equation}
where $\alpha_{il}^{\rm r}\sim {\cal {CN}}(0,1)$ is the corresponding complex Gaussian gain of the $(i,l)$th path. $\psi _{il}^a$ and $\psi_{il}^e$ are the azimuth and elevation AOAs at the RIS, while $\upsilon_{il}^{\rm r}$ denotes the AOD at the UE, respectively.

We first test the convergence behavior of Algorithm 1 and Algorithm 2 for different system setups. It is observed that both Algorithm 1 and Algorithm 2 converge monotonically, which verifies \emph{Proposition 1}. Although as the number of the reflecting elements $N$ grows the convergence speed becomes slower, the average number of external iterations needed to achieve a relatively good performance is only within 100 times even for a large $N$, i.e., $N=256$. Note that while the two algorithms seem to overlap, this does not mean they enjoy the same convergence rate. This only indicates that the number of outer iterations the two algorithms need is similar, and is reasonable since the speed of outer iteration depends on all the four variables, three of which are the same for the two algorithms. As previously stated, the SDR method involves extremely high computational complexity due to a large number of randomization trials while the SCF method enjoys a much lower complexity. Therefore, the time to obtain the optimal $\bm\theta$ for the two methods are totally different. In other words, while Algorithm 1 and Algorithm 2 have similar number of outer iterations, the convergence time for each iteration is different and thus Algorithm 2 still converges faster and has lower computational complexity than Algorithm 1. In other words, Algorithm 2 is capable of striking a satisfactory balance between the system performance and the computational complexity.

\begin{figure}[!t]
\setlength{\abovecaptionskip}{0pt}
\setlength{\belowcaptionskip}{0pt}
\centering %使得插入的照片居中显示
\includegraphics[width=9cm,height=7cm]{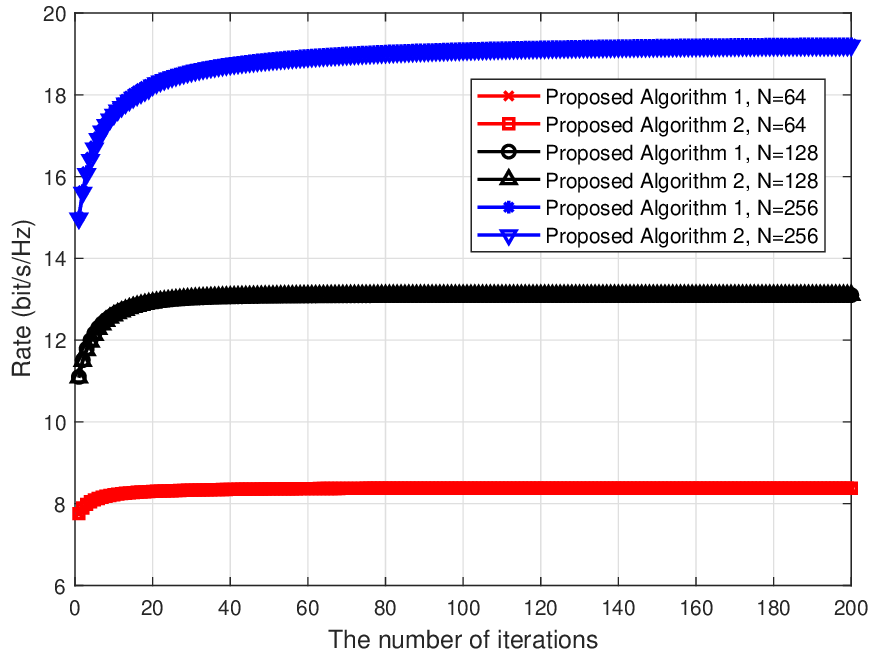} %图片标题
\caption{Convergence behavior of Algorithm 1 and Algorithm 2.}
\label{fig:0}       % 给图片一个标签便于交叉引用
\end{figure}

We compare the proposed Algorithm 1 and Algorithm 2 with the following baseline schemes:
\begin{itemize}
  \item Without RIS: Set ${\bf H}_{\text{eq}}={\bf H}_d$, calculate ${\bf W}_s$ according to (\ref{eq:Ws}).
  \item Random RIS: Randomly generate RIS phase shift and set ${\bf H}_{\text{eq}}$, calculate (\ref{eq:Ws}) according to (\ref{eq:Ws}).
  \item Benchmark 1: An element-wise locally optimal algorithm proposed in \cite{feng2021passive}.
  \item Benchmark 2:  A low-complexity sum-path-gain maximization algorithm proposed in \cite{ning2020beamforming}.
\end{itemize}
\begin{figure}[!t]
\setlength{\abovecaptionskip}{0pt}
\setlength{\belowcaptionskip}{0pt}
\centering %使得插入的照片居中显示
\includegraphics[width=9cm,height=7cm]{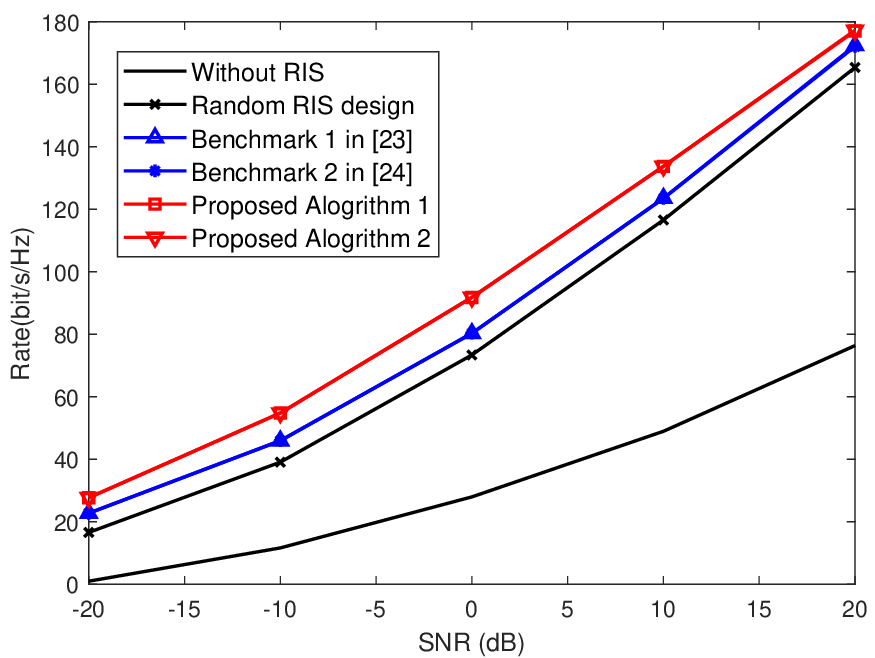} %图片标题
\caption{Achievable rate performance of different schemes versus SNR.}
\label{fig:0}       % 给图片一个标签便于交叉引用
\end{figure}

In Fig. 3 the achievable rate of different schemes as a function of signal-to-noise ratio (SNR) is illustrated. The related parameters are set as $N_t=16,N_r=16,N_s=16$, and $N=120$. First, it can be seen clearly from that all the schemes with RIS outperform that without RIS, even the random RIS beamformig design is much better than the no-RIS case. This is understandable because the channel total gain is significantly improved through RIS, which will be certificated later. Second, the optimized RIS designs, i.e., Benchmark 1, Benchmark 2, Algorithm 1 and Algorithm 2, outperform the random RIS design. Moreover, our proposed algorithms enjoy better performance than Benchmark 1 and Benchmark 2 in all SNRs. This can be illustrated by two reasons. On the one hand, the reformulated problem is equivalent to the original capacity maximization problem in our paper, while in Benchmark 1 and Benchmark 2, the corresponding reformulation is just an approximation on the high SNR region. The two benchmarks perform worse than our proposed algorithms even in high SNR region. On the other hand, our paper has theoretically proven the first-order optimality of our proposed algorithm while Benchmark 1 and Benchmark 2 not. To sum up, our proposed algorithms can better reconfigure the equivalent MIMO channel for transmission by properly adjusting the RIS reflection.
\begin{figure}[!t]
\setlength{\abovecaptionskip}{0pt}
\setlength{\belowcaptionskip}{0pt}
\centering %使得插入的照片居中显示
\includegraphics[width=9cm,height=7cm]{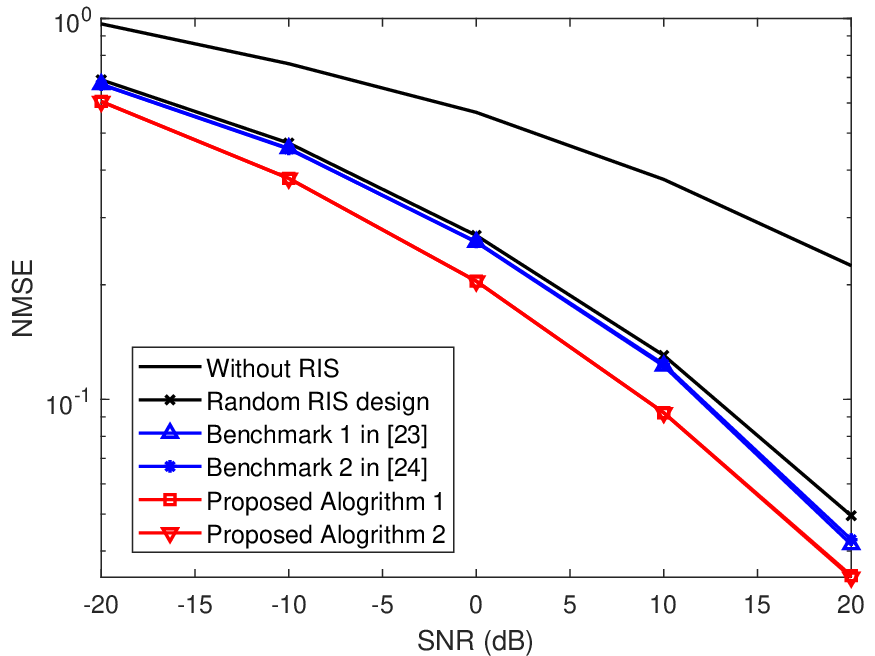} %图片标题
\caption{NMSE performance versus SNR.}
\label{fig:0}       % 给图片一个标签便于交叉引用
\end{figure}
\begin{figure}[!h]
\setlength{\abovecaptionskip}{0pt}
\setlength{\belowcaptionskip}{0pt}
\centering %使得插入的照片居中显示
\includegraphics[width=9cm,height=7cm]{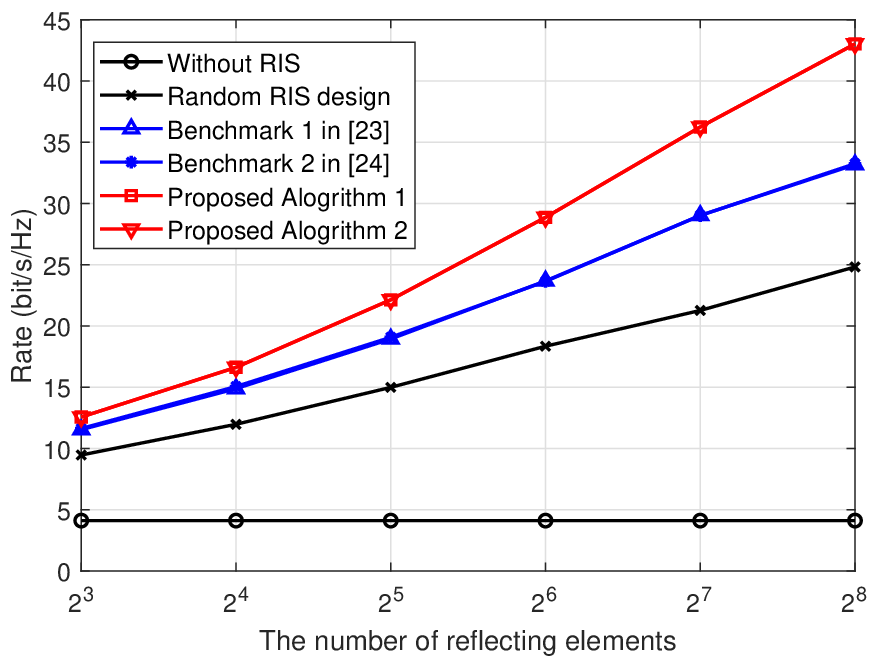} %图片标题
\caption{Achievable rate versus the number of reflecting elements.}
\label{fig:0}       % 给图片一个标签便于交叉引用
\end{figure}

NMSE is another important optimization objective for massive MIMO design problems apart from rate. Fig. 4 illustrates the NMSE curves of various benchmark schemes as a function of SNR. The related parameters used in Fig. 3 are identical to that in Fig. 2. Similar trends can be observed as in Fig. 2 that the two proposed algorithms achieve lower NMSE than other benchmarks, which demonstrates the effectiveness of our proposed algorithms. Besides, it is noted that set the weight matrix ${\bf W}={\bf I}_{N_s}$ will result in the generalized minimum mean squared error design.

We show in Fig. 5 the achievable rate with regard to the number of RIS reflecting elements $N$ at $0$dB with $N_t=4,N_r=4$. The figure shows that the superiority achieved by the proposed designs in comparison to other baseline schemes gradually increases with a growing number of $N$. This is because the joint transceiver and RIS reflector design gains enhanced flexibility in leveraging the spatial degrees of freedom (DoF) introduced by the RIS as $N$ grows, highlighting the advantages of a well-thought-out RIS design strategy.
\begin{figure}[!t]
\setlength{\abovecaptionskip}{0pt}
\setlength{\belowcaptionskip}{0pt}
\centering %使得插入的照片居中显示
\includegraphics[width=9cm,height=7cm]{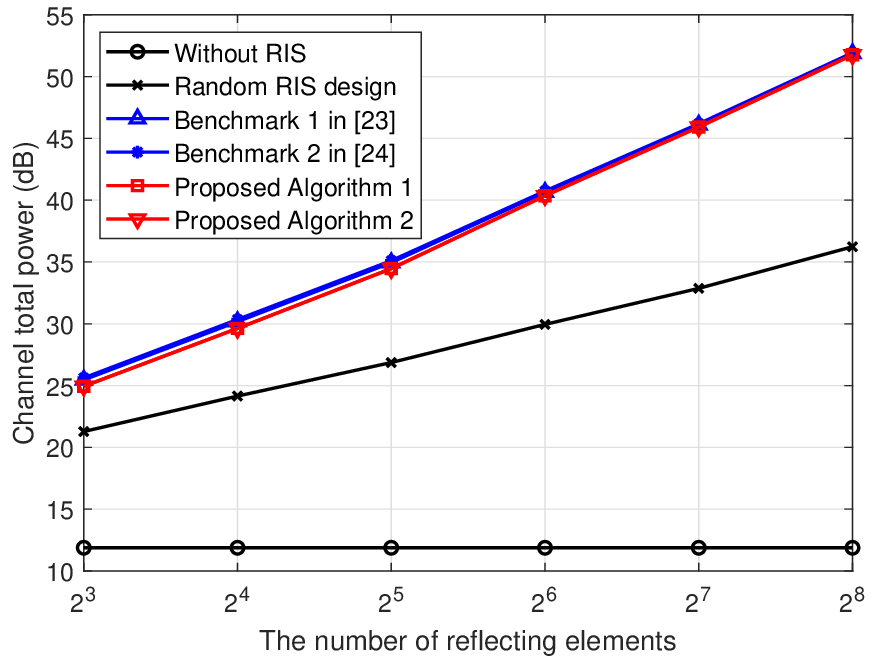} %图片标题
\caption{Channel total power versus the number of reflecting elements.}
\label{fig:0}       % 给图片一个标签便于交叉引用
\end{figure}

In Fig. 6, the channel total power is plotted as a function of the number of RIS reflecting elements $N$ with $N_t=4,N_r=4$ and SNR$=20$dB. It is observed that by deploying and optimizing RIS, the power of the channel is greatly improved. Besides, it is not surprised that Benchmark 1 and Benchmark 2, whose optimization objective are exactly the maximization of channel total power $\|{\bf H}_{\text{eq}}\|_F^2$ beat the other schemes. Furthermore, the channel total powers between the proposed algorithms and the two benchmarks is indistinguishable. On the one hand, it illustrates that the proposed algorithms enjoy a satisfactory performance in terms of the channel energy improvement. On the other hand, it also verifies the feasibility of substituting the channel total power as one alternative design objective in high-SNR region.
\begin{figure}[!t]
\setlength{\abovecaptionskip}{0pt}
\setlength{\belowcaptionskip}{0pt}
\centering %使得插入的照片居中显示
\includegraphics[width=9cm,height=7cm]{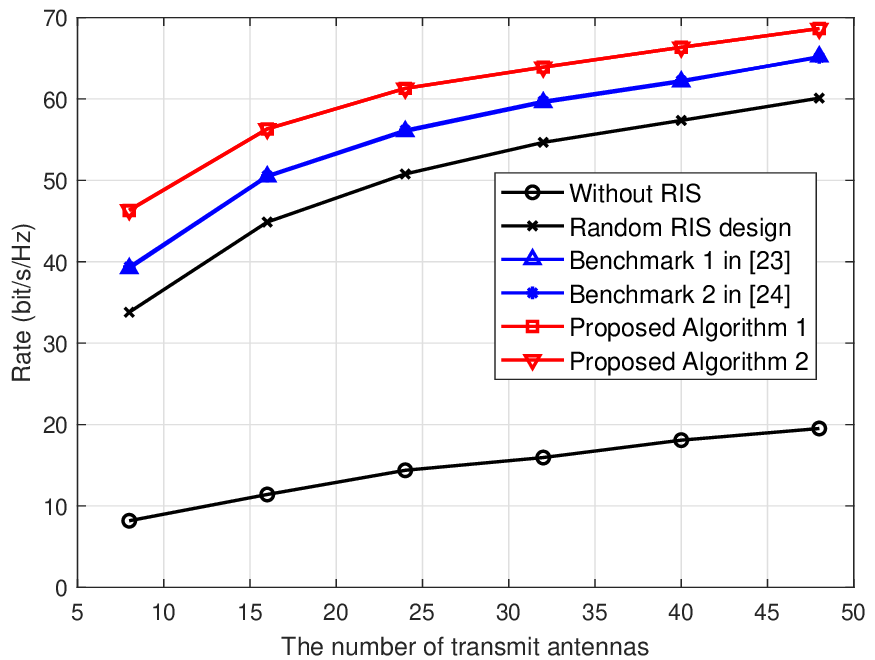} %图片标题
\caption{Achievable rate versus the number of transmit antennas.}
\label{fig:0}       % 给图片一个标签便于交叉引用
\end{figure}

Fig. 7 compares the rate performance of different schemes for different number of transmit antennas $N_t$ with $N_r=8,N=64$ and SNR$=0$dB. It can be observed clearly that when the number of transmit antennas increases, the performance of all schemes improves. This is expected since the transmit beamforming gain increases as the number of transmit antennas increases. Compared with Fig. 5, it is also observed that the rate improvement caused by increasing the number of transmit antennas is minor compared with increasing the number of reflecting elements. Hence, it is more suggested to deploy RIS to assist the MIMO channel communication performance as an energy-efficient substitute of antenna.
\begin{figure}[!t]
\setlength{\abovecaptionskip}{0pt}
\setlength{\belowcaptionskip}{0pt}
\centering %使得插入的照片居中显示
\includegraphics[width=9cm,height=7cm]{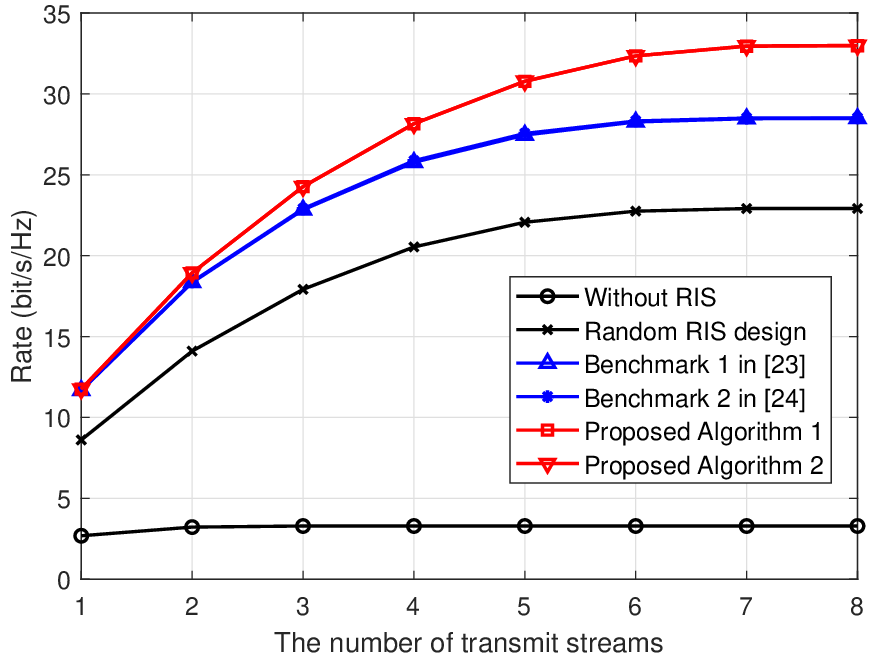} %图片标题
\caption{Achievable rate versus the number of transmit streams.}
\label{fig:0}       % 给图片一个标签便于交叉引用
\end{figure}

Fig. 8 plots the achievable rate comparison of various methods as a function of the number transmit streams $N_s$ with $N_t=16,N_r=16,N=120$ and SNR$=-10$dB. It can be
clearly found that system performance improves as the number of transmit streams $N_s$ grows at first, but when $N_s$ is larger enough, the system performance will saturate. Furthermore, the figure also shows that the performance gap between the proposed algorithms and other RIS design schemes increases as the number of transmit streams increases. This is because when the communication demand is not strong, i.e., $N_s$ is small, the performances gain of the two benchmarks are enough to support the communication requirement, while when the communication demand is intense, the proposed designs that can better extract the DoF promised by RIS are recommended.
\begin{figure}[!t]
\setlength{\abovecaptionskip}{0pt}
\setlength{\belowcaptionskip}{0pt}
\centering %使得插入的照片居中显示
\includegraphics[width=9cm,height=7cm]{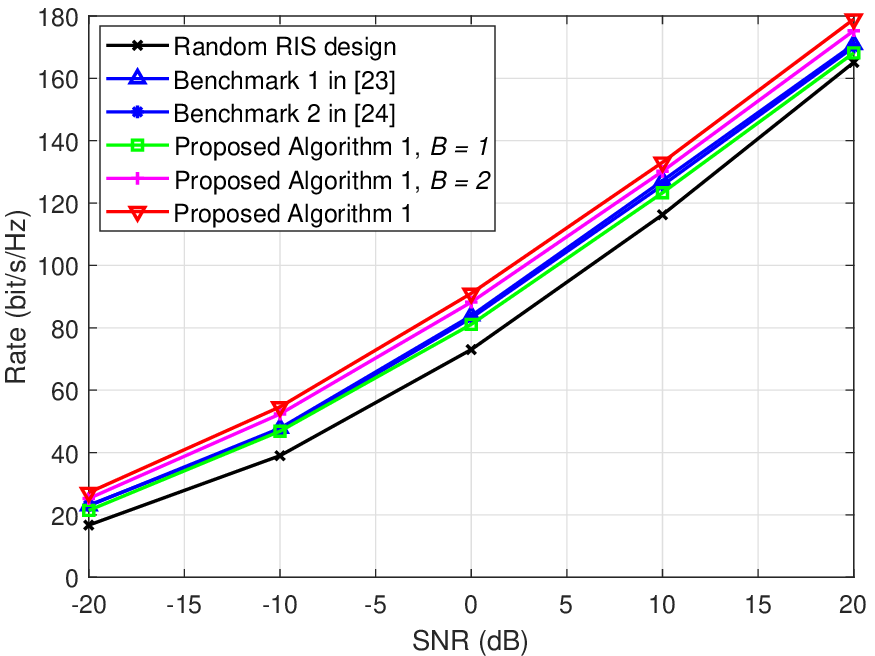} %图片标题
\caption{Rate comparison for Algorithm 1 and its quantized version.}
\label{fig:0}       % 给图片一个标签便于交叉引用
\end{figure}
\begin{figure}[!t]
\setlength{\abovecaptionskip}{0pt}
\setlength{\belowcaptionskip}{0pt}
\centering %使得插入的照片居中显示
\includegraphics[width=9cm,height=7cm]{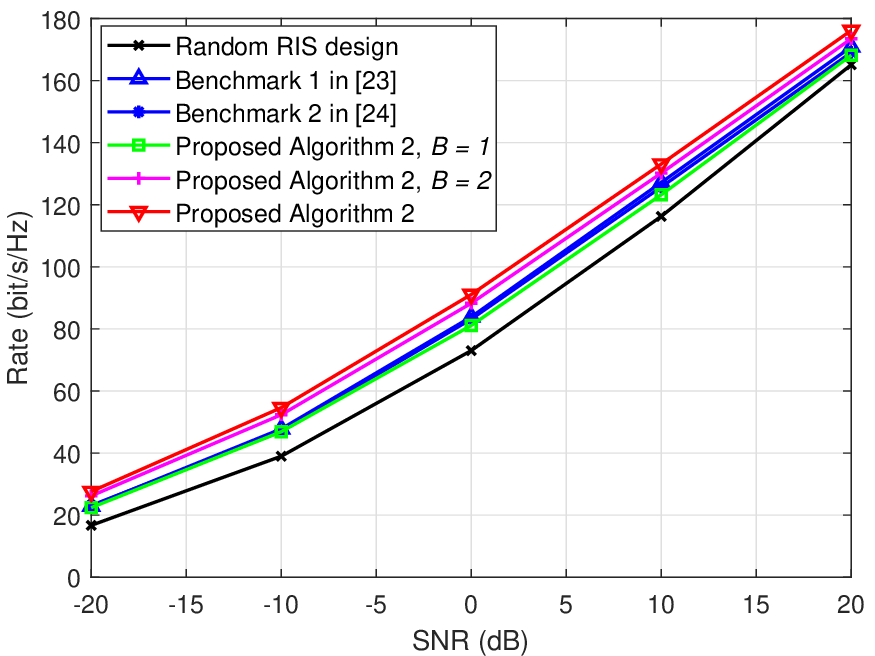} %图片标题
\caption{Rate comparison for Algorithm 2 and its quantized version.}
\label{fig:0}       % 给图片一个标签便于交叉引用
\end{figure}
\begin{figure}[!t]
\setlength{\abovecaptionskip}{0pt}
\setlength{\belowcaptionskip}{0pt}
\centering %使得插入的照片居中显示
\includegraphics[width=9cm,height=7cm]{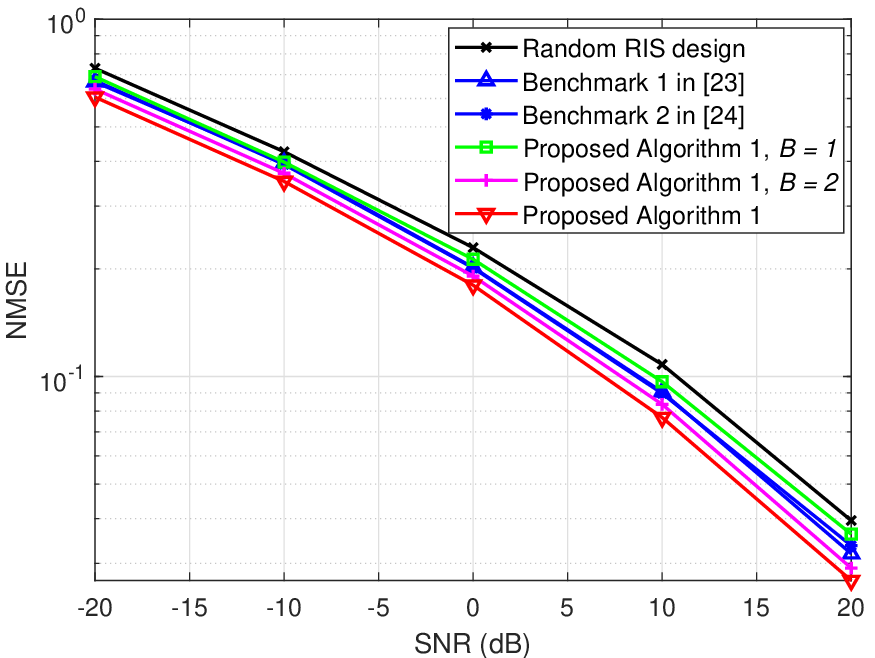} %图片标题
\caption{NMSE comparison for Algorithm 1 and its quantized version.}
\label{fig:0}       % 给图片一个标签便于交叉引用
\end{figure}
\begin{figure}[!t]
\setlength{\abovecaptionskip}{0pt}
\setlength{\belowcaptionskip}{0pt}
\centering %使得插入的照片居中显示
\includegraphics[width=9cm,height=7cm]{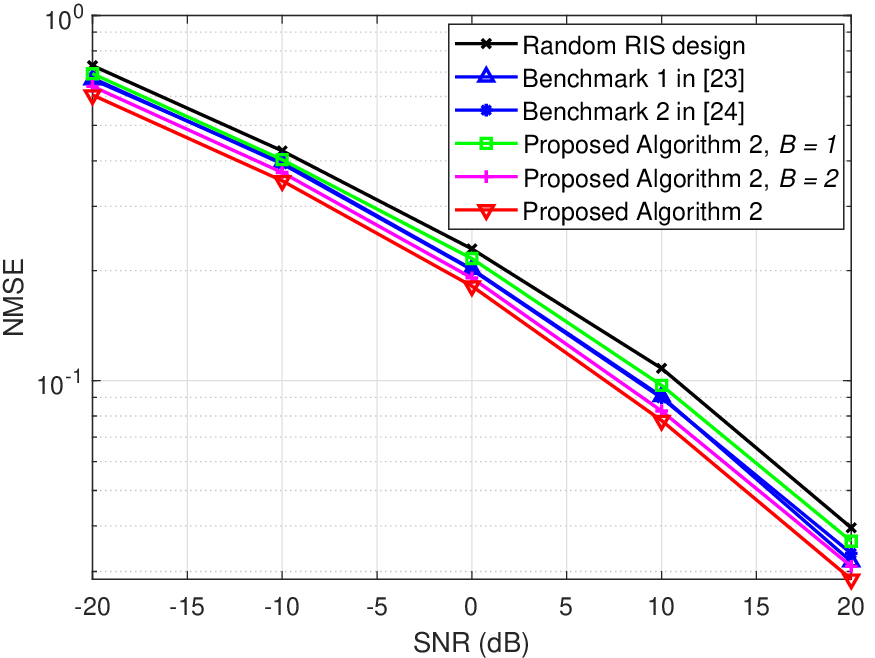} %图片标题
\caption{NMSE comparison for Algorithm 2 and its quantized version.}
\label{fig:0}       % 给图片一个标签便于交叉引用
\end{figure}

Considering the high hardware cost and limited manufacturing capabilities, RIS employs discrete reflections rather than continuous ones in practical systems\footnote{Another effective solution to the hardware cost and energy consumption of massive MIMO is hybrid precoding technique. One efficient way to combine RIS and hybrid precoding is approximating the optimal unconstrained transceiver obtained from the proposed algorithms by using the principle of basis pursuit as in \cite{el2014spatially}.}. Therefore, to assess the effectiveness of our proposed algorithms under these circumstances, we quantize each continuous reflection coefficient to its closest discrete value according to the nearest Euclidean distance. Consequently, the resulting reflection coefficient is $\hat\theta=e^{j\frac{2\pi \hat m}{2^B}}$, with $B$ being the quantization precision and
\begin{equation}\label{eq:discrete}
\widehat m = \arg \mathop {\min }\limits_{m \in \{ 0, 1,\cdots ,{2^B} - 1\} } \left| {\varphi  - \frac{{2\pi m}}{{{2^B}}}} \right|,
\end{equation}
with $\varphi$ being the optimal continuous reflection coefficient calculated in Section \uppercase\expandafter{\romannumeral3}-C.

Next, we show the effect of quantized RIS phase for the proposed algorithms in terms of both rate and NMSE in Fig 9--Fig. 12. The related system parameters are set as $N_t=N_r=16$ and $N=120$. For both Algorithm 1 and Algorithm 2, the system rate increases and the NMSE decreases when the quantization precision bit $B$ increases. Furthermore, the proposed algorithms achieve significant performance gain compared with the two benchmarks even with severe hardware impairments, i.e., $B=2$, which validates its effectiveness. Moreover, it can be observed that 1 bits of quantization precision is enough at high SNR-regime, which indicates that the quantization precision of RIS should be properly chosen according to the specific system parameters to realize a satisfactory balance between the system performance and the hardware implementation since the power consumption and cost of RIS increase with quantization level growth.
\section{Conclusion}
In this paper, joint transceiver and reflection optimization for a RIS-assisted massive MIMO system is investigated. Specifically, we jointly optimize the transmit precoding matrix at the BS, the receive combining matrix at the UE,
and the reflection matrix at
the RIS to maximize the achievable rate. In order to facilitate the mathematical tractability, the non-trivial achievable rate maximization problem was first reformulated into a comparable one, i.e., weighted mean squared error minimization problem. By utilizing the alternating optimization technique, the optimal transmit precoding matrix and the receive combining matrix were derived in
closed forms, and two computationally efficient methods were advocated for the nonconvex RIS reflection optimization problem based on SDR and SCF methods, respectively. The convergence is guaranteed for both SDR-based and SCF-based algorithms. In particular, the SCF-based algorithm is proven to converge to a KKT point of the weighted mean squared error minimization problem. Numerical simulation results validate the superior performance of our proposed algorithms in terms of the achievable rate, the NMSE, and the channel total power.
\begin{appendices}
\section{Proof of Lemma 1}
Denote ${\bf W}_d={\bf C}^{-1}{\bf H}_{\text{eq}}{\bf W}_s$ where ${\bf C}\triangleq {\sigma^2}{\bf I}_{N_r}+{\bf H}_{\text{eq}}{\bf W}_s{\bf W}_s^H{\bf H}_{\text{eq}}^H$. Then by substituting this into (\ref{eq:rate}), we have
\begin{align}\label{eq:rate1}
R&=\text{log}_2\left(\left|{\bf I}_{N_s} + \frac{1}{\sigma ^2}\left({\bf W}_s^H{\bf H}_{\text{eq}}^H{\bf C}^{-1}{\bf C}^{-1}{\bf H}_{\text{eq}}{\bf W}_s\right)^{-1}\right.\right.\nonumber\\
&~\left.\left.\times{\bf W}_s^H{\bf H}_{\text{eq}}^H{\bf C}^{-1}{\bf H}_{\text{eq}}{\bf W}_s{\bf W}_s^H{\bf H}_{\text{eq}}^H{\bf C}^{-1}{\bf H}_{\text{eq}}{\bf W}_s\right|\right)\nonumber\\
&\overset{(\text a)}{=}\text{log}_2\left(\left|{\bf I}_{N_s} +\frac{1}{\sigma ^2}{\bf Q}{\bf V}^{-1}{\bf Q}^H\right|\right),
\end{align}
where (a) utilizes $\text{log}_2|{\bf I}+{\bf AB}|=\text{log}_2|{\bf I}+{\bf BA}|$, and ${\bf Q}\triangleq{\bf W}_s^H{\bf H}_{\text{eq}}^H{\bf C}^{-1}{\bf H}_{\text{eq}}{\bf W}_s, {\bf V}\triangleq {\bf W}_s^H{\bf H}_{\text{eq}}^H{\bf C}^{-2}{\bf H}_{\text{eq}}{\bf W}_s$.

Besides, it is not difficult to obtain the following relationship
\begin{align}\label{eq:QQh}
{\bf Q}^H{\bf Q}&={\bf W}_s^H{\bf H}_{\text{eq}}^H{\bf C}^{-1}{\bf H}_{\text{eq}}{\bf W}_s{\bf W}_s^H{\bf H}_{\text{eq}}^H{\bf C}^{-1}{\bf H}_{\text{eq}}{\bf W}_s\nonumber\\
&={\bf W}_s^H{\bf H}_{\text{eq}}^H{\bf C}^{-1}({\bf C}-\sigma^2{\bf I}_{N_r}){\bf C}^{-1}{\bf H}_{\text{eq}}{\bf W}_s\nonumber\\
&={\bf W}_s^H{\bf H}_{\text{eq}}^H{\bf C}^{-1}{\bf H}_{\text{eq}}{\bf W}_s-\sigma^2{\bf W}_s^H{\bf H}_{\text{eq}}^H{\bf C}^{-2}{\bf H}_{\text{eq}}{\bf W}_s\nonumber\\
&={\bf Q}-\sigma^2{\bf V}.
\end{align}

According to (\ref{eq:QQh}), we have ${\bf V}=\frac{{\bf Q}-{\bf Q}^H{\bf Q}}{\sigma^2}$. By plugging this into (\ref{eq:rate1}), it yields
\begin{align}\label{eq:relation}
R&=\text{log}_2\left(\left|{\bf I}_{N_s} +{\bf Q}\left({\bf Q}-{\bf Q}^H{\bf Q}\right)^{-1}{\bf Q}^H\right|\right)\nonumber\\
&\overset{(\text a)}{=}\text{log}_2\left(\left|\left({\bf I}_{N_s} -{\bf Q}^H\right)^{-1}\right|\right)\nonumber\\
&=\text{log}_2\left|{\bf E}^{-1}\right|,
\end{align}
where (a) is obtained by using the matrix inversion lemma.

Equipped with (\ref{eq:relation}), \emph{Lemma 1} is then proven by following similar derivations as in \cite{shi2011iteratively}.
\section{Proof of Lemma 2}
Denote the objective function and feasible set of $({\cal{QP}}^{(i)})$ by $q({\bf x}^{(i)})$ and ${\cal F}_{i-1}$, respectively. Let ${\bf x}_{(i-1)}=\left[{\text{Re}}\{{\bm\theta}_{(i-1)}\}^T,{\text{Im}}\{{\bm\theta}_{(i-1)}\}^T,1\right]^T$. Since problem $({\cal{QP}}^{(i)})$ is convex and ${\bf x}_{(i-1)}\in{\cal F}_{i}$, then we have
\begin{equation}\label{eq:xn}
q({\bf x}_{(i-1)})\geq q({\bf x}^{(i)}).
\end{equation}
Define
\begin{small}
\begin{equation}\label{eq:gtheta}
t({\bm\theta})={\bm\theta}^H\left({\bf A}_r^H{\bf A}_r+\lambda{\bf I}_{2N+1}\right){\bm\theta}\!-\!2\text{Re}\left\{{\bm\theta}^H{\bf A}_r^H{\bf a}_r\right\}+{\bf a}_r^H{\bf a}_r+\lambda.
\end{equation}
\end{small}
Since $t({\bm\theta})$ can be regarded as the equivalent complex-valued form of $q(\bf x)$, it is not difficult to obtain the following relationship according to (\ref{eq:xn})
\begin{equation}\label{eq:gn_relation}
t({\bm\theta}_{(i-1)})\geq t({\bm\theta}^{(i)}).
\end{equation}
Recall that ${\bm\theta}_{(i)}$ is the unit-modulus version of ${\bm\theta}^{(i)}$, we can decompose ${\bm\theta}^{(i)}$ as
\begin{equation}\label{eq:composition}
{\bm\theta}^{(i)}={\bm\theta}_{(i)}+{\bf M}{\bm\theta}_{(i)},
\end{equation}
where ${\bf M}=\text{diag}[m_1,m_2,\cdots,m_N]$ and $m_n=|\theta_n^{(i)}|-1\geq 0$ with $\theta_n^{(i)}$ being the $n$th element of ${\bm\theta}^{(i)}$. Here we utilize the fact that $|\theta_n^{(i)}|\geq 1,\forall n\in\{1,2,\cdots,N\}$. Then we have
\begin{align}\label{eq:gn_relation1}
&t\left({\bm\theta}^{(i)}\right)-t\left({\bm\theta}_{(i)}\right)\nonumber\\
&\!\!\overset{\text{(a)}}=\!\!2 \lambda {\bm\theta}_{(i)}^H {\bf M}{\bm\theta}_{(i)}\!\!+\!{\bm\theta}_{(i)}^H\left({\bf A}_r^H{\bf A}_r{\bf M}\!+\!{\bf M} {\bf A}_r^H{\bf A}_r\right){\bm\theta}_{(i)}\!\!-\!\!{\bm\theta}_{(i)}^H {\bf M}{\bf A}_r^H{\bf a}_r \nonumber\\
&~+{\bm\theta}_{(i)}^H {\bf M}\left({\bf A}_r^H{\bf A}_r+\lambda {\bf I}_{2N+1}\right){\bf M}{\bm\theta}_{(i)}-{\bf a}_r^H{\bf A}_r{\bf M} {\bm\theta}_{(i)}\nonumber\\
&~\geq 2 \lambda{\bm\theta}_{(i)}^H {\bf M} {\bm\theta}_{(i)}+{\bm\theta}_{(i)}^H\left({\bf A}_r^H{\bf A}_r{\bf M}+{\bf M} {\bf A}_r^H{\bf A}_r\right) {\bm\theta}_{(i)}\nonumber\\
&~~-{\bf a}_r^H{\bf A}_r{\bf M}{\bm\theta}_{(i)}-{\bm\theta}_{(i)}^H {\bf M}{\bf A}_r^H{\bf a}_r \nonumber\\
&~\overset{\text{(b)}}= 2 \lambda\left\|{\bf M} {\bm\theta}_{(i)}\right\|_1+{\bm\theta}_{(i)}^H\left({\bf A}_r^H{\bf A}_r{\bf M}+{\bf M} {\bf A}_r^H{\bf A}_r\right) {\bm\theta}_{(i)}\nonumber\\
&~~-{\bf a}_r^H{\bf A}_r{\bf M}{\bm\theta}_{(i)}-{\bm\theta}_{(i)}^H{\bf M}{\bf A}_r^H{\bf a}_r,
\end{align}
where (a) holds since ${\bf A}_r^H{\bf A}_r+\lambda {\bf I}_{2N+1}$ is positive definite and (b) holds because ${\bm\theta}_{(i)}^H {\bf M} {\bm\theta}_{(i)}=\sum\limits_{n = 1}^N m_n|{\theta_n}_{(i)}|^2 =\sum\limits_{n = 1}^N |m_n{\theta_n}_{(i)}|=\left\|{\bf M} {\bm\theta}_{(i)}\right\|_1$. Based on [\cite{zhang2011matrix}, Theorem 7.5], we further obtain
\begin{align}\label{eq:lambda_relation}
&{\bm\theta}_{(i)}^H\left({\bf A}_r^H{\bf A}_r{\bf M}+{\bf M} {\bf A}_r^H{\bf A}_r\right) {\bm\theta}_{(i)}\geq -\frac{N}{4}\lambda_{\text{max}}({\bf A}_r^H{\bf A}_r)\lambda_{\text{max}}({\bf M})\nonumber\\
&\overset{\text{(a)}}=-\frac{N}{4}\lambda_{\text{max}}({\bf A}_r^H{\bf A}_r)\left\|{\bf M} {\bm\theta}_{(i)}\right\|_{\infty},
\end{align}
where (a) is obtained by utilizing the fact that $\left\|{\bf M} {\bm\theta}_{(i)}\right\|_{\infty}=\text{max}_n|m_n{\theta_n}_{(i)}|=\lambda_{\text{max}}({\bf M})$.
By substituting (\ref{eq:lambda_relation}) into (\ref{eq:gn_relation1}), it yields
\begin{small}
\begin{align}\label{eq:gn_relation2}
&t\left({\bm\theta}^{(i)}\right)-t\left({\bm\theta}_{(i)}\right)\nonumber\\
& \geq 2 \lambda\left\|{\bf M} {\bm\theta}_{(i)}\right\|_1-\frac{N}{4}\lambda_{\text{max}}({\bf A}_r^H{\bf A}_r)\left\|{\bf M} {\bm\theta}_{(i)}\right\|_{\infty}\nonumber\\
&~-{\bf a}_r^H{\bf A}_r{\bf M}{\bm\theta}_{(i)}-{\bm\theta}_{(i)}^H{\bf M}{\bf A}_r^H{\bf a}_r\nonumber\\
& \geq 2 \lambda\left\|{\bf M} {\bm\theta}_{(i)}\right\|_1-\frac{N}{4}\lambda_{\text{max}}({\bf A}_r^H{\bf A}_r)\left\|{\bf M} {\bm\theta}_{(i)}\right\|_1\nonumber\\
&~-{\bf a}_r^H{\bf A}_r{\bf M}{\bm\theta}_{(i)}-{\bm\theta}_{(i)}^H{\bf M}{\bf A}_r^H{\bf a}_r\nonumber\\
& \geq 2 \lambda\left\|{\bf M} {\bm\theta}_{(i)}\right\|_1\!\!-\!\!\frac{N}{4}\lambda_{\text{max}}({\bf A}_r^H{\bf A}_r)\left\|{\bf M} {\bm\theta}_{(i)}\right\|_1\!\!-\!\!2\left\|{\bf A}_r^H{\bf a}_r\right\|_2\left\|{\bf M}{\bm\theta}_{(i)}\right\|_2 \nonumber\\
& \geq 2\lambda\left\|{\bf M} {\bm\theta}_{(i)}\right\|_1-\left(\frac{N}{4} \lambda_{\text{max}}({\bf A}_r^H{\bf A}_r)+2\left\|{\bf A}_r^H{\bf a}_r\right\|_2\right)\left\|{\bf M}{\bm\theta}_{(i)}\right\|_1 \nonumber\\
& \geq 0,
\end{align}
\end{small}
where $\left\|{\bf M}{\bm\theta}_{(i)}\right\|_1\geq \left\|{\bf M}{\bm\theta}_{(i)}\right\|_{\infty},\left\|{\bf A}_r^H{\bf a}_r\right\|_2\left\|{\bf M}{\bm\theta}_{(i)}\right\|_2\geq {\bm\theta}_{(i)}^H{\bf M}{\bf A}_r^H{\bf a}_r, \left\|{\bf M}{\bm\theta}_{(i)}\right\|_1\geq \left\|{\bf M}{\bm\theta}_{(i)}\right\|_2$ and $\lambda\geq \frac{N}{8}\lambda_{\text{max}}({\bf A}_r^H{\bf A}_r)+\|{\bf A}_r^H{\bf a}_r\|_2$ are respectively utilized in deriving the last four inequalities.

Combining (\ref{eq:gn_relation2}) and (\ref{eq:gn_relation}), it gives
\begin{equation}\label{eq:gn_relation4}
t({\bm\theta}_{(i-1)})\geq t({\bm\theta}_{(i)})
\end{equation}
Denote the objective function of (\ref{eq:problem6}) as $h(\bm\theta)$. Since ${\bm\theta}_{(i)}^H{\bm\theta}_{(i)}={\bm\theta}_{(i-1)}^H{\bm\theta}_{(i-1)}=N$, it is easy to obtain
\begin{equation}\label{eq:fn_relation}
h({\bm\theta}_{(i-1)})-h({\bm\theta}_{(i)})=t({\bm\theta}_{(i-1)})-t({\bm\theta}_{(i)})\geq 0.
\end{equation}
Consequently, the sequences $\{h({\bm\theta}_{(i)})\}_{i=0}^{\infty}$ is non-increasing. On the other hand, since the objective of (\ref{eq:problem6})  has a finite lower bound, e.g., $h(\bm\theta)\geq 0$. Therefore, it finally converges to a finite value.

Assume that the SCF mathod converges at the $I$th iteration, then we prove that $|{\theta}^{(i)}_n|=1,\forall n\in\{1,2,\cdots,N\}$ holds when $i\geq I$. Obviously, we can obtain the following expression upon convergence
\begin{equation}\label{eq:fn_convergence}
h({\bm\theta}_{(i-1)})-h({\bm\theta}_{(i)})=t({\bm\theta}_{(i-1)})-t({\bm\theta}_{(i)})\ \approx  0.
\end{equation}
Combining (\ref{eq:gn_relation2}) and (\ref{eq:gn_relation1}), it can be readily proven that
\begin{align}\label{eq:middle}
&2 \lambda{\bm\theta}_{(i)}^H {\bf M} {\bm\theta}_{(i)}+{\bm\theta}_{(i)}^H\left({\bf A}_r^H{\bf A}_r{\bf M}+{\bf M} {\bf A}_r^H{\bf A}_r\right) {\bm\theta}_{(i)}\nonumber\\
&-{\bf a}_r^H{\bf A}_r{\bf M}{\bm\theta}_{(i)}-{\bm\theta}_{(i)}^H {\bf M}{\bf A}_r^H{\bf a}_r\geq 0,
\end{align}
when $\lambda\geq \frac{N}{8}\lambda_{\text{max}}({\bf A}_r^H{\bf A}_r)+\|{\bf A}_r^H{\bf a}_r\|_2$. Then we have
\begin{align}\label{eq:convergence}
0&=t\left({\bm\theta}^{(i)}\right)-t\left({\bm\theta}_{(i)}\right)\nonumber\\
&= 2 \lambda {\bm\theta}_{(i)}^H {\bf M}{\bm\theta}_{(i)}+{\bm\theta}_{(i)}^H\left({\bf A}_r^H{\bf A}_r{\bf M}+{\bf M} {\bf A}_r^H{\bf A}_r\right){\bm\theta}_{(i)}\nonumber\\
&\!+{\bm\theta}_{(i)}^H {\bf M}\left({\bf A}_r^H{\bf A}_r\!\!+\!\!\lambda {\bf I}_{2N+1}\right){\bf M}{\bm\theta}_{(i)}\!\!-\!\!{\bf a}_r^H{\bf A}_r{\bf M} {\bm\theta}_{(i)}\!\!-\!\!{\bm\theta}_{(i)}^H {\bf M}{\bf A}_r^H{\bf a}_r\nonumber\\
&~\overset{\text{(a)}}\geq {\bm\theta}_{(i)}^H {\bf M}\left({\bf A}_r^H{\bf A}_r+\lambda {\bf I}_{2N+1}\right){\bf M}{\bm\theta}_{(i)}\nonumber\\
&~\geq\lambda{\bm\theta}_{(i)}^H {\bf M}^2{\bm\theta}_{(i)}\nonumber\\
&=\lambda\sum\limits_{n = 1}^N m_n^2,
\end{align}
where (a) is obtained by utilizing (\ref{eq:middle}). Recall the definition of ${\bm\theta}^{(i)}$ in (\ref{eq:composition}), since $\lambda\geq 0$, so (\ref{eq:convergence}) holds if and only if ${\bf M}={\bf 0}$, which further indicates that ${\bm\theta}^{(i)}$ satisfies the unit-modulus constraint.

Considering the proof of the optimality condition for (\ref{eq:problem6}), since (\ref{eq:problem8}) is the real-valued version of (\ref{eq:problem6}), it is sufficient to prove that ${\bf x}^{(i)}$ meets the KKT conditions for (\ref{eq:problem6}). The KKT conditions for problem (\ref{eq:problem8}) are listed as follows
\begin{align}\label{eq:KKT1}
&2({\bf R}+\lambda {\bf I}_{2N+1}){\bf x}+\sum_{n=1}^{N+1} 2 \eta_n {\bf E}_n {\bf x} ={\bf 0}, \nonumber\\
&{\bf x}^{T}{\bf E}_n{\bf x}=1, \quad n \in\{1,2, \cdots, N+1\}
\end{align}
where $\bm\eta$ is the corresponding Lagrange multiplier of (\ref{eq:problem8}) and $\eta_n$ is the $n$th element of $\bm\eta$.

Recall that ${\bf x}^{(i)}$ is the global optimal solution for the convex problem $({\cal{QP}}^{(i)})$, it must satisfies the KKT conditions for $({\cal{QP}}^{(i)})$, which gives
\begin{align}\label{eq:KKT2}
&2({\bf R}+\lambda {\bf I}_{2N+1}){\bf x}^{(i)}+{{\bf B}^{(i)}}^T\overline{\bm\eta}^{(i)}={\bf 0}, \nonumber\\
&{\bf B}^{(i)}{\bf x}^{(i)}={\bf 1},
\end{align}
where $\overline{\bm\eta}^{(i)}$ is the Lagrange multiplier of $({\cal{QP}}^{(i)})$.

The second equation in (\ref{eq:KKT2}) yields
\begin{align}\label{eq:cosine1}
&\cos \left(\text{arg} \left\{\theta_n^{(i-1)}\right\}\right) \text{Re}\left\{\theta_n^{(i)}\right\}+\nonumber\\
&\sin \left(\text{arg} \left\{\theta_n^{(i-1)}\right\}\right) \text{Im}\left\{\theta_n^{(i)}\right\}\!=\!1,\forall n\in\{1,2,\cdots, N\}
\end{align}
Since ${\bm\theta}^{(i)}$ is the unit-modulus solution, (\ref{eq:cosine1}) is rewritten as
\begin{equation}\label{eq:cosine2}
\!\!\!\!\!\cos \left(\text{arg}\left\{\!\theta_n^{(i-1)}\!\right\}\!-\!\text{arg} \left\{\theta_n^{(i)}\right\}\right)\!=1,\forall n\in\{1,2,\cdots, N\}
\end{equation}
which indicates that
\begin{equation}\label{eq:thetan}
\theta_n^{(i)}=e^{\text{arg} \left\{\theta_n^{(i-1)}\right\}}.\quad\forall n\in\{1,2,\cdots, N\}
\end{equation}
Therefore, it is easy to obtain
\begin{align}\label{eq:bn}
&{\bf E}_n{\bf x}^{(i)}={\bf b}_n^{(i)},\\
&{{\bf x}^{(i)}}^{T}{\bf E}_n{\bf x}^{(i)}=1, \quad n \in\{1,2, \cdots, N+1\}
\end{align}
where ${\bf b}_n^{(i)}$ is the $n$th column of ${{\bf B}^{(i)}}^T$.

Substituting (\ref{eq:bn}) into (\ref{eq:KKT1}), we have
\begin{align}\label{eq:KKT3}
&2({\bf R}+\lambda {\bf I}_{2N+1}){\bf x}^{(i)}+\sum_{n=1}^{N+1} \overline{\eta}_n^{(i)} {\bf E}_n {\bf x}^{(i)} ={\bf 0}, \nonumber\\
&{{\bf x}^{(i)}}^{T}{\bf E}_n{\bf x}^{(i)}=1, \quad n \in\{1,2, \cdots, N+1\}
\end{align}
where $\overline{\eta}_n^{(i)}$ is the $n$th element of $\overline{\bm\eta}^{(i)}$. Comparing (\ref{eq:KKT3}) with (\ref{eq:KKT1}), it can be found that the converged solution to problem ${\cal{QP}}^{(i)}$, i.e., ${\bf x}^{(i)}$ exactly satisfies the KKT conditions of problem (\ref{eq:problem8}) with ${\bm\eta}=\frac{1}{2}\overline{\bm\eta}^{(i)}$.

Together, \emph{Lemma 2} is proven that the SCF-based RIS solution is able to converge to a KKT point of the optimization problem (\ref{eq:problem8}).
\end{appendices}
\bibliographystyle{IEEEtran}
\bibliography{reference}
\end{document}